\documentclass[twocolumn]{aastex63}

\accepted{August 24, 2022}
\shorttitle{On the formation of over-ionized plasma in evolved supernova remnants}
\shortauthors{Katsuragawa et al}

\begin{document}

\title{On the formation of over-ionized plasma in evolved supernova remnants}

\correspondingauthor{Miho Katsuragawa}
\email{miho.katsuragawa@ipmu.jp}

\author{Miho Katsuragawa}
\affiliation{Kavli Institute for the Physics and Mathematics of the Universe (WPI), The University of Tokyo, Kashiwa 277-8583, Japan}

\author{Shiu-Hang Lee}
\affiliation{Department of Astronomy, Kyoto University, Kitashirakawa Oiwake-cho, Sakyo-ku, Kyoto-shi, Kyoto 606-8502, Japan}
\affiliation{Kavli Institute for the Physics and Mathematics of the Universe (WPI), The University of Tokyo, Kashiwa 277-8583, Japan}

\author{Hirokazu Odaka}
\affiliation{Department of Physics, The University of Tokyo, Hongo, Bunkyo-ku, Tokyo 113-0033, Japan }
\affiliation{Kavli Institute for the Physics and Mathematics of the Universe (WPI), The University of Tokyo, Kashiwa 277-8583, Japan}

\author{Aya Bamba}
\affiliation{Department of Physics, The University of Tokyo, Hongo, Bunkyo-ku, Tokyo 113-0033, Japan }
\affiliation{Research Center for the Early Universe, School of Science, The University of Tokyo, 7-3-1 Hongo, Bunkyo-ku, Tokyo 113-0033, Japan}

\author{Hideaki Matsumura}
\affiliation{Kavli Institute for the Physics and Mathematics of the Universe (WPI), The University of Tokyo, Kashiwa 277-8583, Japan}

\author{Tadayuki Takahashi}
\affiliation{Kavli Institute for the Physics and Mathematics of the Universe (WPI), The University of Tokyo, Kashiwa 277-8583, Japan}
\affiliation{Department of Physics, The University of Tokyo, Hongo, Bunkyo-ku, Tokyo 113-0033, Japan }

%========================= Abstract =========================
\begin{abstract}
One of the outstanding mysteries surrounding the rich diversity found in supernova remnants (SNRs) is the recent discovery of over-ionized or recombining plasma from a number of dynamically evolved objects.
To help decipher its formation mechanism, we have developed a new simulation framework capable of modeling the time evolution of the ionization state of the plasma in an SNR.
The platform is based on a one-dimensional hydrodynamics code coupled to a fully time-dependent non-equilibrium ionization calculation, accompanied by a spectral synthesis code to generate space-resolved broadband X-ray spectra for SNRs at arbitrary ages.
We perform a comprehensive parametric survey to investigate the effects of different circumstellar environments on the ionization state evolution in SNRs up to a few 10$^4$ years. 
A two-dimensional parameter space, spanned by arrays of interstellar medium (ISM) densities and mass-loss rates of the progenitor, is used to create a grid of models for the surrounding environment, in which a core-collapse explosion is triggered.
Our results show that a recombining plasma can be successfully reproduced in case of a young SNR (a few 100 to 1,000 years old) expanding fast in a spatially extended low-density wind, an old SNR ($>$ a few 1,000~years) expanding in a dense ISM, or an old SNR broken out from a confined dense wind region into a tenuous ISM.
Finally, our models are confronted with observations of evolved SNRs, and an overall good agreement is found except for a couple of outliers. 
\end{abstract}


%=======================================================
%===================== Introduction ====================
\section{Introduction} 

Recent X-ray observations have revealed that some supernova remnants (SNRs) show pieces of evidence of ``over-ionized" or recombining plasma in which the ionization states of heavy elements are found to be higher than their collisional ionization equilibrium (CIE) values \citep[e.g.][]{kawasaki2002, troja2008, ozawa2009, yamaguchi2009, miceli2010, ohnishi2011, ergin2017, katsuragawa2018}. Those evidences of recombining plasma have been typically found in dynamically evolved mixed-morphology SNRs (MM-SNRs) which have bright radio shells and center-filled X-ray emitting regions dominated by thermal emission \citep{RandP1998}. Before its discovery, recombining plasma had been viewed as an unexpected phenomenon under the standard picture of SNR evolution, for that the ionization timescale $\sim$10$^5$~yr in the interstellar medium (ISM) with a density $\sim$ 0.1--1 cm$^{-3}$ is typically longer than the age of SNRs, hence the shock-heated plasma in SNRs was always expected to be either in an ionizing state or approaching CIE. The presence of recombining plasma hence indicates a non-trivial evolution history unaccountable by the standard picture, but its formation mechanism is still not fully understood.

In general, the dynamical and ionization evolutions in SNRs are highly related to their circumstellar medium (CSM) with which the interactions can contribute importantly to the rich diversity found so far in the morphologies and emission properties of known SNRs. In particular, since the properties of the X-ray emitting plasma in SNRs are sensitive to the surrounding medium, studying the origin of SNR recombining plasma can in turn provide us a powerful tool for understanding this diversity. Currently there exists two main scenarios proposed to explain the occurrence of RP, both involve a rapid decrease of the electron temperature $T_e$. One proposed a fast rarefaction of the hot plasma when the SNR blastwave breaks out from a dense CSM into a lower density region \citep{IandM1989, yamaguchi2018}. Most of the MM-SNRs showing recombining plasma are found to be consistent with a core-collapse (CC) origin, so the dense CSM can possibly be produced by a pre-supernova (SN) stellar wind with an enhanced episodic mass loss. When the shock breaks out from the dense CSM, the electron temperature drops quickly due to adiabatic expansion. 
In the other scenario, interaction with dense and clumpy molecular clouds, which is found to be the case for many MM-SNRs, can lead to a fast cooling of the electrons via the evaporation of the dense cold molecular clumps and their thermal conduction with the surrounding hot plasma \citep{kawasaki2002, matsumura2017a, okon2018}.

Essentially, the cooling timescale of the electrons becomes shorter than the recombination timescales of the heavy ions, leading to an over-ionization state. While both the scenarios appear to be able to account for the occurrence of RP, self-consistent numerical models which can describe both the long-term hydrodynamic and ionization-state evolutions in an SNR are still scarce, especially in the context of understanding the origin of recombining plasma.
We therefore need to solve the time evolution of the electron temperature and the ionization fractions of all abundant elements as local variables in a framework of hydrodynamics.
\textbf{In this study, we in particular focus on unraveling the link between this evolution and the diverse circumstellar environments associated with CCSNRs using a systematic survey on a grid of one-dimensional (1D) hydrodynamic models, while reserving a multi-dimensional approach for future work.} Such a 1D model allows us to investigate the influence of the wind and surrounding gas structure on the temperature and ionization evolutions of SNRs, producing the parameter trends in various environment conditions.

In this work, we have constructed a numerical framework that takes into account both the hydrodynamics and ionization state simultaneously based on a 1D Lagrangian hydrodynamics code. Using this code, we performed a comprehensive parameter survey to quantitatively assess the effect of various CSM environments on the ionization states in SNR plasma in a fully time-dependent fashion. To confront observations, we couple our hydrodynamic simulation with a spectral synthesis code that can compute X-ray continua and line emissions in accordance with the electron temperature and ion fraction profiles calculated by our simulations and the latest atomic database. In particular, we explore the model dependence of the electron temperature ($T_e$) and ionization temperature ($T_Z$) which can be directly compared to those extracted from analyses of observed X-ray spectra. Some previous works have attempted to approach similar tasks \citep[e.g.][]{IandF1984, IandM1989, ellison2007, patnaude2009, ellison2010, lee2014, zhang2019}, but their applications have mainly focused on the study of young SNRs which do not show any sign of recombining plasma \citep[e.g.,][]{lee2013, slane2014}. 
Our new framework enables us to simulate SNR evolution up to an age $> 10^4$ years after the transition to the radiative phase, and predict the long-term evolution of the X-ray spectra which can apply to old dynamically evolved SNRs. This framework is developed upon an existing SNR modeling platform \textit{CR-hydro-NEI} \citep[e.g.,][and reference therein]{patnaude2009, lee2012, lee2014}, which has been previously applied to studying the X-ray evolution of young CC \citep{jacovich2021} and type Ia SNRs \citep{martinez2018} successfully. Specifically, our code includes physical processes of shock heating, temperature equilibration through Coulomb interactions, radiative cooling, non-equilibrium evolution of ionization states of various heavy ions and so on, fully coupled to a hydrodynamic simulation without post-processing.

This paper is organized as follows. Section \ref{sec:model} provides a description of the methodology including the main components of our models; Section \ref{sec:exa} discusses the results from a fiducial model to showcase the capabilities of our code; Section \ref{sec:results} focuses on a parameter survey to assess the dependence of our models on different CSM environments, and compares our results with current observations. In Section \ref{sec:disc}, we discuss possible formation mechanisms of recombining plasma in SNRs and limitation of the present model employing 1D hydrodynamics. Section \ref{sec:conc} gives the conclusions of this work.

%========================= Model =========================
\section{Model}\label{sec:model}

In order to study the evolution of plasma including recombining plasma in SNRs, we develop a new model to calculate the X-ray emission that can follow time evolution of plasma in SNRs for a few $10^{4}$~yr. We follow the evolution using a Lagrangian hydrodynamics code VH-1 \citep[e.g.,][]{BE2001} in which each grid memorizes the chemical, thermal, and ionization conditions of a mass element. We assume a spherical symmetry for the SNR for simplicity, considering only the radial variation of key physical properties, as summarized in Table \ref{tab:valiable}.
The simulations include an implementation of full time and spatially dependent non-equilibrium ionization (NEI) to keep track of the evolution of ionization states of various chemical elements in both the SN ejecta and the shocked ambient medium, using which we synthesize the broadband X-ray spectrum at any selected SNR age and radial position.
In the following sections, we will explain the methodology of our calculations using an example of an SNR model evolved from a CC-SN progenitor and fiducial parameters.

\begin{deluxetable}{lc}
	%\tablenum{1}
	\tablecaption{Key variables in our simulation.\label{tab:valiable}}
	\tablewidth{0pt}
	\tablehead{
		\colhead{Variable}      & \colhead{Symbol}
	}
	\startdata
		Radius                  & $r(i, t)$          \\
		Gas density                 & $\rho(i, t)$       \\
		Gas velocity                & $v(i, t)$          \\
		Gas pressure                & $P(i, t)$          \\
		Ion densities             & $n_\mathrm{ion}^j(i, t)$ \\
		Electron density        & $n_{e}(i, t)$      \\
		Ion temperatures         & $T_\mathrm{ion}^j(i, t)$ \\
		Electron temperature    & $T_e(i, t)$     \\
	\enddata
	\tablecomments{$i$ and $j$ are the indices for the mass grid number and ion species, and $t$ is the SNR age.}
\end{deluxetable}

%=================== Physical processes ===================
\subsection{Physical processes}

\subsubsection{Temperature evaluation}

Modeling the X-ray emission from the plasma in SNRs requires the temperature evolution of electrons and ions being evaluated self-consistently with the hydrodynamic evolution of the remnants.
We consider the cooling and heating processes of particles in a plasma, including collisionless shock heating, adiabatic compression and expansion, Coulomb collisions between particles with different temperatures and masses, and radiative cooling.
We will describe these processes in the following.

Radiative cooling can be an important effect in optically thin plasma usually found in the shocked plasma in SNRs, especially for electron temperatures in the range of $10^4$--$10^7$~K. Radiative cooling is accounted for in every mass grid in our models,
so that as a particle loses energy through radiation, the total energy loss (erg s$^{-1}$) is described by
\begin{eqnarray}
	\frac{dE}{dt} = - \int n_Hn_e\Lambda_N(T)dV,
\end{eqnarray}
where $\Lambda_N$ is the cooling rate in units of erg s$^{-1}$ cm$^{3}$ \citep{schure2009}.
The overall cooling rates are calculated by summing over the contributions of all different elements,
for which the densities are calculated using the local elemental abundances and ion fractions as follows:
\begin{eqnarray}
	\Lambda_N(T) = \sum_{Z} \frac{n_Z}{n_Z(solar)}\Lambda_N(Z, T).
\end{eqnarray}
We adopt the cooling rate $\Lambda_N(Z, T)$ from \citet{schure2009} under the assumption that plasma is in a CIE state for simplicity, although we do follow the NEI of the plasma which in principle can be used for synthesizing the partial cooling rates and hence the cooling curve at each grid and time step.
We postpone such a more consistent calculation to future works.
The CIE assumption for the cooling rate would result in less effective radiative cooling in an over-ionized state, but this does not have a large impact for investigating the parameter trend associated with the surrounding environment.

Another important physical process is the interaction between electrons and ions through Coulomb collisions because their temperatures are typically out of equilibrium in SNR plasma.
The free electrons and ions in each grid are allowed to exchange energy under the local Coulomb timescale and their temperatures are updated accordingly.

Assuming there are an equilibrating particle $a$ and its target background particle $b$ with temperatures $T_a$ and $T_b$, the Coulomb equilibration timescale between $a$ and $b$ is calculated by \citet{spitzer1962} as
\begin{eqnarray}
	\tau_\mathrm{eq}(a, b) = \frac{3}{8\sqrt{2\pi}} \frac{m_am_b}{n_bZ_a^2Z_b^2e^4{\rm log}\lambda} \left(\frac{k_BT_a}{m_a} + \frac{k_BT_b}{m_b}\right)^{\frac{3}{2}},
\end{eqnarray}
where $m$, $Z$, $e$ and $k_B$ denote the particle mass, the atomic number, the elementary charge, and the Boltzmann constant, respectively;
$\mathrm{log}\lambda=\mathrm{log}(b_\mathrm{max}/b_\mathrm{min})$ is the so-called Coulomb logarithm for the collisional pair, and $b_\mathrm{max}$ and $b_\mathrm{min}$ denote the upper and lower limits of the collision parameter between the particles.
In the plasma, $b_\mathrm{max}$ is given by the Debye length as $b_\mathrm{max} = \sqrt{kT_e/(4\pi e^2n_e)}$, and $b_\mathrm{min} = \mathrm{max}(b^\mathrm{classical}_\mathrm{min}, b^\mathrm{quantum}_\mathrm{min}$) \citep[e.g.,][]{callen2006}.
For electron–ion collisions, $b^\mathrm{classical}_\mathrm{min}= Z_be^2/(3kT_e)$ and $b^\mathrm{quantum}_\mathrm{min} = h/(4\pi\sqrt{3kT_em_e})$.
For the case of ion–ion collisions, $b^\mathrm{classical}_\mathrm{min} = Z_aZ_be^2/(m_\mathrm{cm}\bar{u}^2)$ and $b^\mathrm{qm}_\mathrm{min} = h/(4\pi m_{cm}\sqrt{\bar{u}^2})$, where the mass $m_\mathrm{cm}$ in the center-of-mass system is defined by the reduced mass, $m_am_b/(m_a + m_b)$.
The average of $\bar{u}^2$ over the distribution of target particles is $\bar{u}^2 = 3(kT_a/m_a + kT_b/m_b)$.
When a particle interacts in dense gas, the temperatures of the various particle species can become close in a relatively short time compared to the dynamical time scale since the equilibration timescale is inversely proportional to the density of the gas.
The temperature change of the particles $a$ is calculated by 
\begin{equation}
\frac{dT_a}{dt} = \frac{T_b - T_a}{t_\mathrm{eq}}.
\end{equation}

\subsubsection{Non-equilibrium ionization and recombination}
The shape of the emitted X-ray spectra at any given SNR age depends on the electron temperature and the electron and ion densities.
Instead of evaluating these quantities in a post-processing manner as in some previous works in the literature, we calculate the time evolution of the ionization state for all elements included in our models along with the hydrodynamic evolution.
In our simulations, the ion fractions of 13 elements are estimated based on their ionization and recombination rates.
The rates of ionization $I_j$ and recombination $R_j$ of an atom per unit time is given by
 \begin{eqnarray}
	 I_j = C_{{\rm ion}, j}(T_e)n_e,\ \ \ R_j = C_{{\rm rec}, j} (T_e)n_e,
\end{eqnarray}
where $j$ is the ionization charge number, and $C_{{\rm ion}, j} (T)$ and $C_{{\rm rec}, j} (T)$ are the ionization and recombination coefficients, respectively, which depend on the electron temperature $T_e$.
We employ the values of these rates from \citet{patnaude2009} and reference therein.
The ion fraction X$^{t+1}_{j}$ in each Lagrangian gas element is calculated by
\begin{eqnarray}
	{\rm X}^{t+1}_j =& &dtI_{j-1}{\rm X}^t_{j-1} \nonumber \\
	& & + (1 - dt(I_j + R_j)){\rm X}^t_j +dtR_{j+1}{\rm X}^t_{j+1},
\end{eqnarray}
where $X(t+dt)$ means the next time step.
The first term on the right-hand side refers to the ionization from state $j-1$, and the last term to the recombination from the ion state $j+1$.
The second term refers to the fractional decrease of ions with state $j$ after a time step $dt$.
We adopt an implicit Crank-Nicolson scheme to ensure numerical stability. Like the temperatures, these ion fractions are also traced in each mass coordinate throughout the SNR lifetime. 

%================ Initial conditions ===================
\subsection{Initial conditions}\label{sec:init_cond}
\begin{figure}[tb]
	\centering
	\includegraphics[width=80mm]{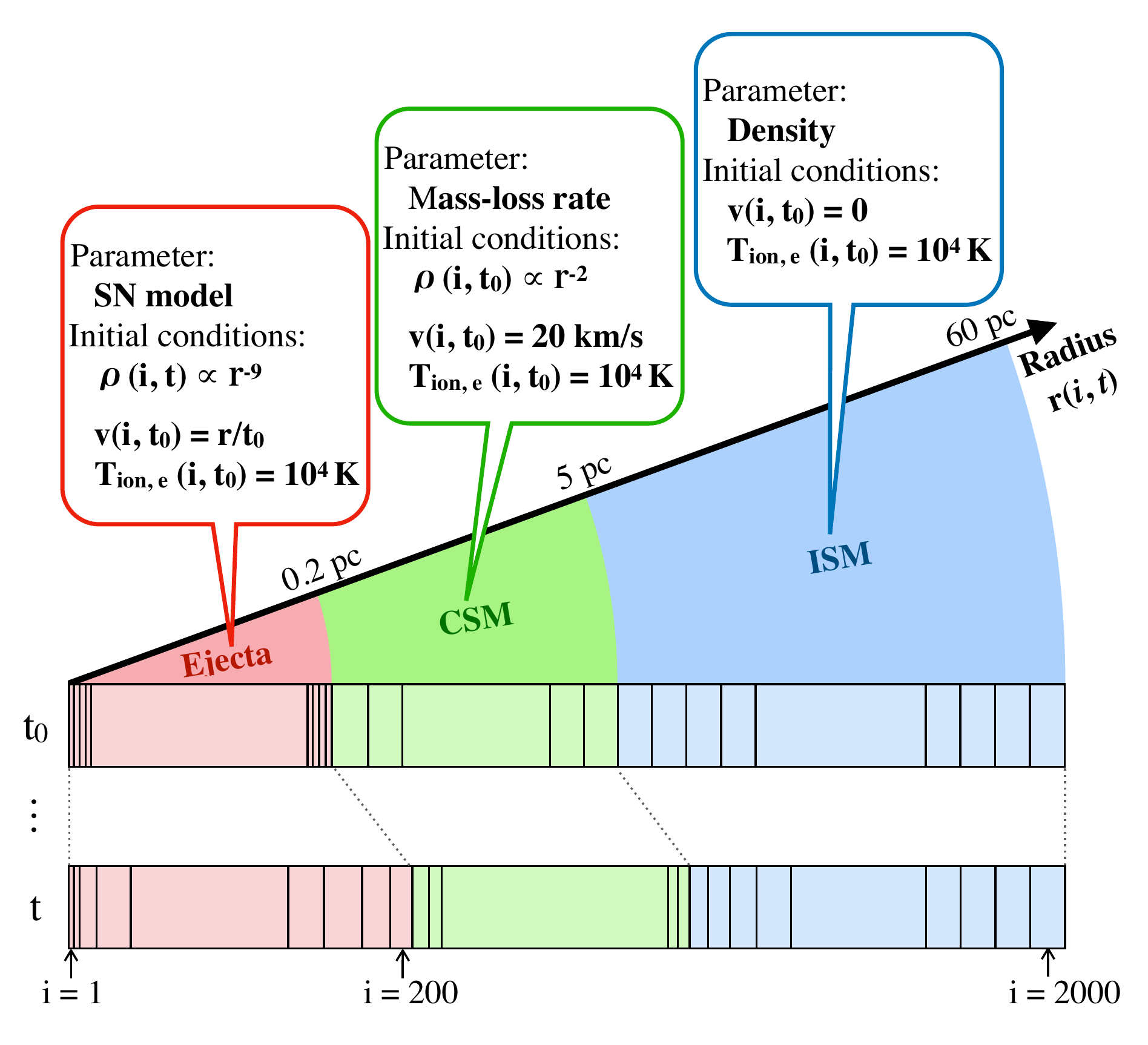}
	\includegraphics[width=60mm]{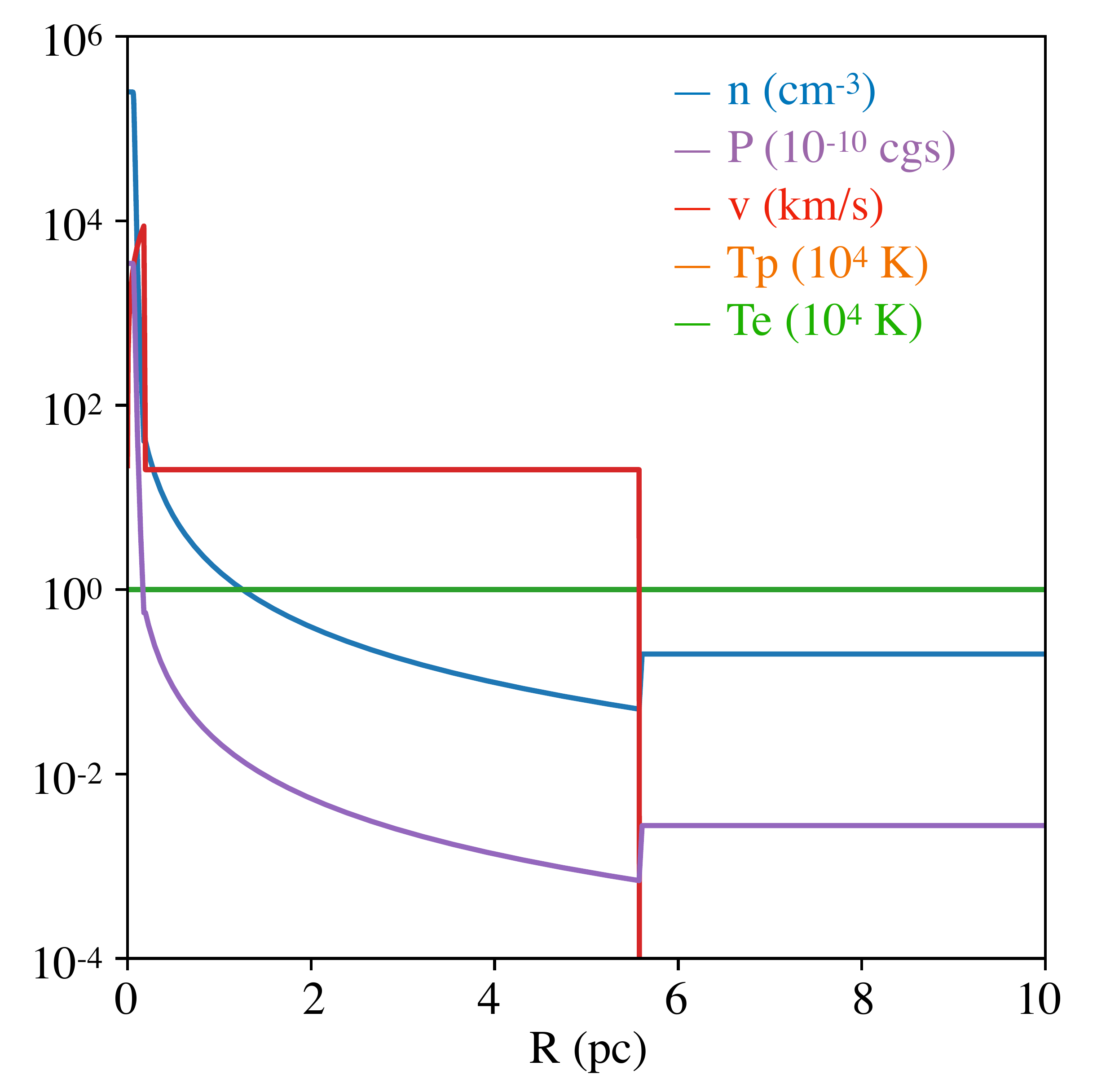}
	\caption{Top panel: a schematic diagram illustrating the simulation setup for a fiducial model. The horizontal bar shows the distribution of Lagrangian grids assigned to the three main regions in a model: the SN ejecta, CSM created by a pre-SN stellar wind, and the ambient ISM. Bottom panel: initial conditions for the fiducial model. The proton ($T_p$) and electron ($T_e$) temperatures are set to be in equilibrium at $10^4$~K everywhere initially. The initial temperatures of ions ($T_{\rm ion}$) are same as the electron temperature.}
	\label{fig:s25_init}
\end{figure}

In this study, we consider a model for the circumstellar medium (CSM) of an SNR with two main components, each parameterized by the mass-loss rate of the SN progenitor and the density of the surrounding ISM, respectively. The model is composed of an SN ejecta, a wind-like CSM, and a uniform ISM as shown in Fig.~\ref{fig:s25_init} (top). The total number of mass layers for each calculation is fixed at 2000, and ten percent of the grids is assigned to the ejecta.

Our models adopt initial conditions for the gas density, pressure, velocity, temperatures and chemical abundances assuming a free expansion of the SNR ejecta up to 20 yr after the explosion, which is the starting time of the simulations.
A typical initial configuration is shown in Fig.~\ref{fig:s25_init} (bottom).
In this paper, particle acceleration through diffusive shock acceleration is not considered.

For the ejecta, we use a power-law density profile ($\rho_{\rm ejecta} \propto r^{-n}$) with a plateau in the inner core (Chevalier 1982, Truelove \& Mckee 1999).
In the ejecta, densities  of free electrons $n_{\rm e}(i, t_0)$ and ions $n_{\rm ion}(i, t_0, Z)$ are determined according to the abundance distribution of the SN model.
The initial temperatures of electron and ions, $T_{\rm e}(i, t_0)$ and $T_{\rm ion}(i, t_0, Z)$, in the ejecta are assumed to be 10$^{4}$~K at which X-ray radiation is not emitted.
Pressure $P(i, t_0)$ is calculated from $P(i, t_0)$ = $\rho k_BT(i, t_0)$.

\begin{figure}[t]
	\centering
	\includegraphics[width=80mm]{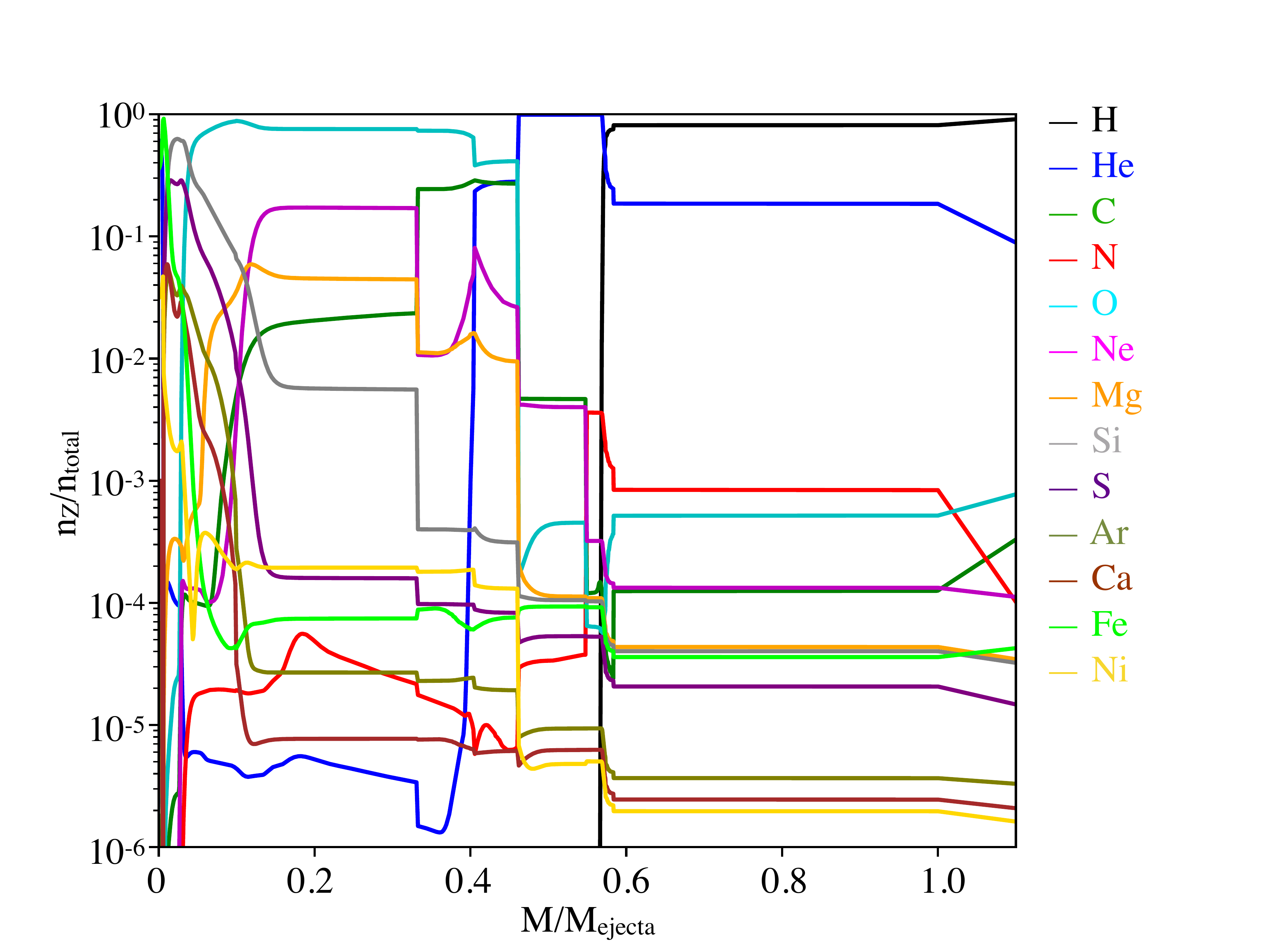}
	\caption{Chemical composition of model s25D as functions of the enclosed mass. The values at $\rm M/M_{ejecta}$ = 1.1 show the ISM abundance.}
	\label{fig:s25abd}
\end{figure}

For the chemical abundances in the ejecta, we make use of a CC SN model \textit{s25D} from a massive red supergiant (RSG) progenitor with a stellar-mass (at the zero-age main sequence) of 25~$M_{\odot}$ based on \citet{RandT2001}.
Fig.~\ref{fig:s25abd} shows the abundance distribution of the \textit{s25D} model.
The initial kinetic energy $E_{\rm SN}$ and the mass of the ejecta are $1.2 \times 10^{51}$~erg and 12.3~$M_{\odot}$, respectively.
We assume that all elements in the unshocked regions are 10\% singly ionized at $t=t_0$ to facilitate the initiation of NEI after shock-heating (e.g., ${\rm X_{C\ I}}$, ${\rm X_{C\ II}}$, ${\rm X_{C\ III}}$, ..., ${\rm X_{C\ VII}}$ = 0.9, 0.1, 0, ..., 0). This treatment practically prevents the calculation from breaking down due to zero free electron density. Similar initial conditions were employed by \citet{patnaude2009}.
The choice of the initial ionization state does not affect our results in any significant way, since the initial state is sufficiently low ionized (almost neutral) and the plasma immediately gets highly ionized when shocked.

We assume that the CSM densities is $\rho_{\rm CSM} = \dot{M}(4\pi v_w)^{-1} r^{-2}$ corresponding to the pre-SN wind \citep{chevalier1982, CandL1994}, 
where $\dot{M}$ is the mass-loss rate of the progenitor and $v_w$ is the velocity of the stellar wind.
To assess the effect of shock interaction with the CSM of different densities on the X-ray emission in the SNR phase, we treat the mass-loss rate of the progenitor as a primary control parameter, but the total mass in the CSM from the RSG wind is fixed at 8~$M_{\odot}$ according to the \textit{s25D} evolution model.
We only consider a steady wind in this study so that $\dot{M}$ is time-independent for simplicity, but our simulation code itself is capable of treating time-dependent mass loss as well. 
The wind-like CSM is smoothly connected to the outer homogeneous ISM region with a constant density.
In other words, we neglect a small shell-like structure generated around the contact between the wind and the ISM in order to investigate the time evolution of the global structure of an SNR. This simplification has little effect on the result of the global hydrodynamics.

The ISM density is another parameter of our models. The contact discontinuity (CD) between the wind and ISM is at a radial position where the sum of the CSM mass reaches the total mass loss of 8~$M_{\odot}$, but the exact location depends on the mass-loss rate.
We performed an independent pure hydro simulation for each $\dot{M}$ and ISM density to follow the evolution of the CSM structure by steadily injecting wind material into a uniform ISM, from which the CD location can be traced until CC explosion.
If the mass-loss rate of the progenitor is relatively small, the density of the CSM is small compared to the ISM near the CD which resembles an embedded low-density bubble situation.
In these cases, the CSM is connected to the ISM at a radius where a mechanical equilibrium between the ram pressure of the wind and the pressure in the ISM is satisfied.
For higher mass-loss rates, a high-density CSM surrounds the vicinity of the ejecta with the outer radius dictated mainly by the time duration of the mass loss instead.
The velocity of the steady wind is fixed at 20~km~s$^{-1}$.
For the elemental abundances in the CSM and ISM, we assume cosmic abundances as in \citet{Wilms2000}, shown in Fig.~\ref{fig:s25abd}.

%=========================================================

%=============== Result of a typical case ================

\section{Simulation outputs - an example}\label{sec:exa}
In this section, we first demonstrate the capabilities of our simulation framework by elaborating on the results from a fiducial model setup.
For the fiducial model, we assume a mass-loss rate of $1\times 10^{-5}\ M_{\odot}\ yr^{-1}$ and an ISM density of 0.2~$\rm cm^{-3}$ as typical values for RSG progenitors.% and from SNR observations.

\subsection{Time evolution} \label{sec:time_evo}
\begin{figure}[tb]
	\centering
	\includegraphics[width=70mm]{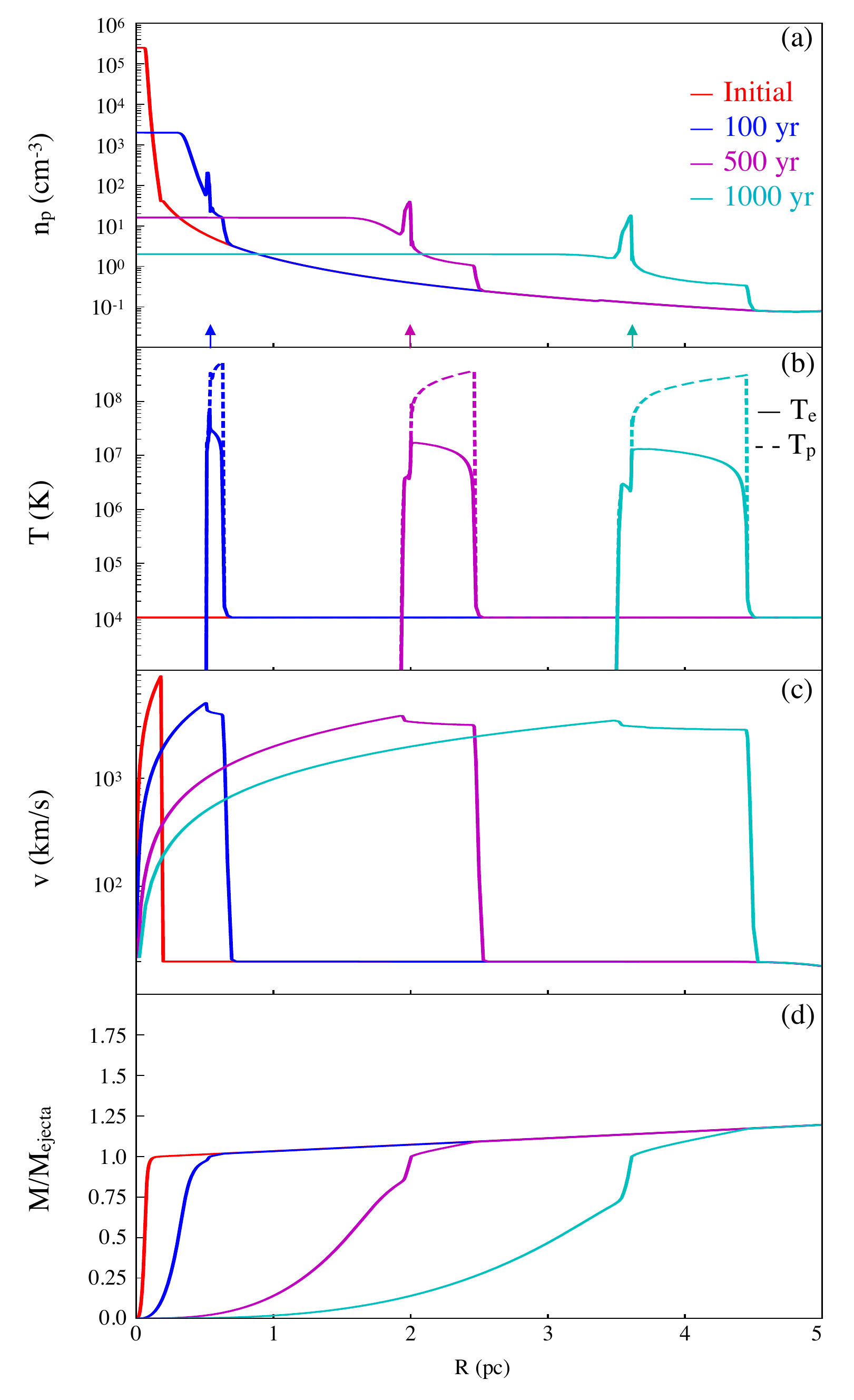}
	\caption{Time evolution of the radial profiles of (a) total gas number density , (b) electron and proton temperatures, (c) gas velocity, and (d) accumulated mass normalized by the ejecta mass for the fiducial model (see text). Arrows in the panel (a) shows the contact discontinuity.}
	\label{fig:timeevol}
\end{figure}

We first present the time evolution of the plasma properties,  focusing on the gas density and electron temperature, which are the most crucial for calculating X-ray emissivities.
Our model provides properties in each fluid element at arbitrary ages.
Examples of hydrodynamic profiles are presented in Fig.~\ref{fig:timeevol} as time snapshots.
Two shockwaves (forward and reverse shocks) are clearly visible and separated by the CD between the ejecta and CSM at 100 years (panel (a)).
The forward shock (FS) propagates in the CSM and reaches around r = 4.5~pc from the explosion center at the age of 1000~yr.
At the same age, the reverse shock (RS) is heating up the ejecta at around r = 3.5~pc.
For a typical RSG wind as in this model, the expansion of the FS (thus SNR size) is fast with an FS velocity staying at around 3000~$\rm km\ s^{-1}$ after 1000~yr of expansion and sweeping up wind material (panel (c)).
The FS has not reached the uniform ISM region yet at this age, after which the FS is expected to decelerate at a much higher rate.

Accordingly, the RS is still weak and moves slowly inward in the rest frame of the unshocked ejecta, because the shocked CSM mass is still relatively small compared to the ejecta mass, as shown in panel (d).
The shocks and subsequent Coulomb interactions heat electrons up to temperatures of $ 10^{6}$--$10^{7}$~K (panel (b)).
After collisionless shock heating, the temperature ratios between the electrons and ions are assumed to be in proportion to their mass ratios, but the ratios typically become smaller gradually due to the energy exchange mediated by their mutual Coulomb interactions.
In the dense region around the CD, the electron temperature nearly equals the proton temperature since the timescale of the Coulomb interaction is shorter than the dynamical and radiative cooling timescales.

\begin{figure}[t]
	\centering
	\includegraphics[width=70mm]{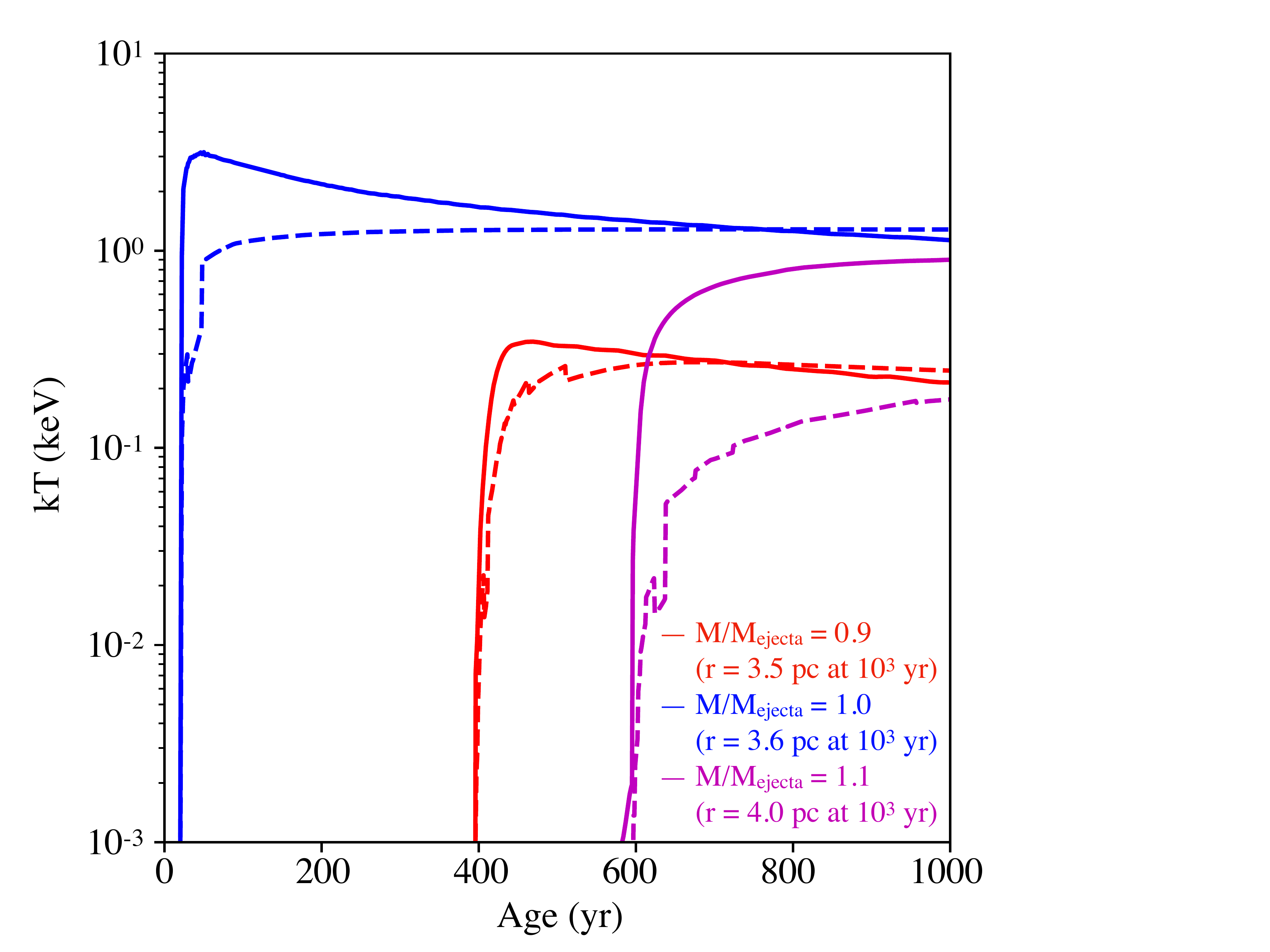}
	\caption{Time evolution of $T_e$ (solid) and $T_Z$ (dashed) for fluid elements located in the ejecta (red), CD (blue), and CSM (magenta).}
	\label{fig:timeevol_te}
\end{figure}

To compare our results effectively with X-ray observations, we define the ionization temperature ($T_Z$), which is a commonly used indicator for gauging the ionization degrees of an X-ray emitting plasma in observational analyses.
$T_Z$ of plasma is defined as an electron temperature of an imaginary plasma in its CIE state that would produce the same ion fractions in the plasma of interest.
$T_Z$ mildly varies with different chemical elements responsible for the X-ray line emission because of their different ionization potentials.
In this paper, we adopt sulphur (S) as our standard since X-ray line emissions from sulphur have been detected in many evolved SNRs including IC~443, a prototypical example for MM-SNRs exhibiting a radiative recombination continuum (RRC) as spectroscopic evidence of recombining plasma.
In addition, $T_Z$ calculated using sulphur has a good sensitivity for a temperature range (0.2--2 keV) close to those obtained from observations \citep{kawasaki2005}.
We calculate $T_Z$ using the number ratios between the most abundantly populated ion state and its neighboring states in sulphur and their comparisons with the expected ratios in a CIE plasma with the same $T_e$.

Fig.~\ref{fig:timeevol_te} shows the time evolution of $T_e$ and $T_Z$ at three mass coordinates in the shocked ejecta, at the CD, and in the shocked CSM for the fiducial model.
The electron temperature in the ejecta (solid red line) rapidly increases by Coulomb interaction and reaches equilibrium with the protons shortly after the shock heating, because the timescale of Coulomb interaction is shorter than the dynamical time scale.
After that it makes a transition to a decreasing phase due to the fast adiabatic expansion of the shocked ejecta. 
The RS decelerates very rapidly inside the dense unshocked ejecta in the early phase, and stays weak in the first 1000~yr.  

In the outer region corresponding to the shocked CSM (magenta lines), $T_e$ keeps increasing gradually via Coulomb heating by the protons.
This is because of the lower average gas densities in the wind, so that it does not reach full equilibrium with the proton temperature in a short timescale like in the ejecta.
$T_Z/T_e$ also approaches unity (CIE) much more slowly in the CSM because the averaged ionization timescale proportional to the density squared is long compared to the fast expansion.
Here, $T_Z$ is always lower than $T_e$, indicating that the shocked CSM is in an ionizing state throughout the first 1000 years of evolution.

On the other hand, in the denser regions around the CD and shocked ejecta, $T_e$ rises almost momentarily to its peak. 
The ionization process also advances quickly in the dense plasma and $T_Z$ goes up with $T_e$ within a few ten years immediately after the shock heating. 
When the density becomes thin by the expansion, the recombination timescale becomes longer than the dynamical timescale, so that the evolution of $T_Z$ lags behind $T_e$ in this phase.
Therefore, the plasma changes from an ionizing state ($T_z < T_e$) to a recombining state ($T_Z > T_e$).
This transition in terms of ionization balance seems to be universal for SNR ejecta expanding inside a tenuous steady wind, as we will further explain in the following sections \citep{moriya2012}.

\begin{figure}[t]
	\centering
	\includegraphics[width=70mm]{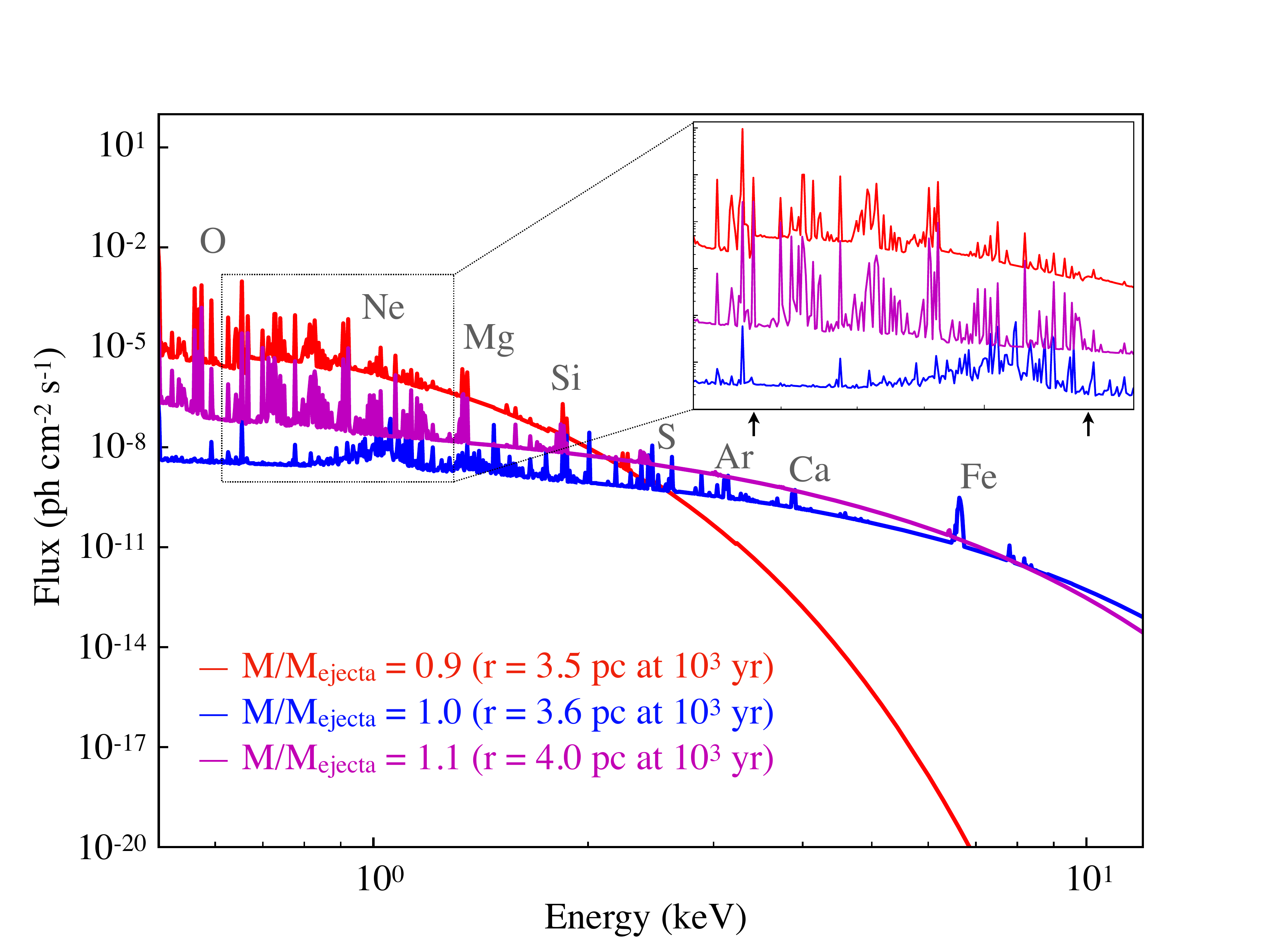}
	\caption{Synthesized X-ray spectra from three locations ($\rm M/M_{\rm ej}$ = 0.9, 1.0, and 1.1) at an age of 1000~yr for the fiducial model. Arrows in the upper right panel shows the edge of RRCs around 0.65~keV and 1.2~keV.}
	\label{fig:spec_5}
\end{figure}
%================ Spectra synthesis ======================
 \subsection{Spectral synthesis}
 
Our spectrum generator provides spatially resolved X-ray spectra at arbitrary ages extracted from the output of the simulation.
We use the {\it AtomDB} database (version 3.0.9) \citep{foster2012} for atomic data needed for calculating X-ray emission from hot, collisionally ionized plasma with $10^4 \le T_e \le 10^9$~K.
{\it AtomDB} provides two FITS files containing a list of lines and continua which have information on the wavelengths, emissivities, and related properties for various ions of different elements, and they are grouped by the electron temperature. 
Thus, we can obtain the energies and emissivities of all lines emitted from an ion at any given electron temperature. 
Our code requires as input the radius of a Lagrangian grid, grid width, electron temperature, densities of electrons and ions, and ion fractions of each element, which are obtained from the output file of the hydrodynamics code.

\begin{figure*}[t]
	\centering
	\includegraphics[width=70mm]{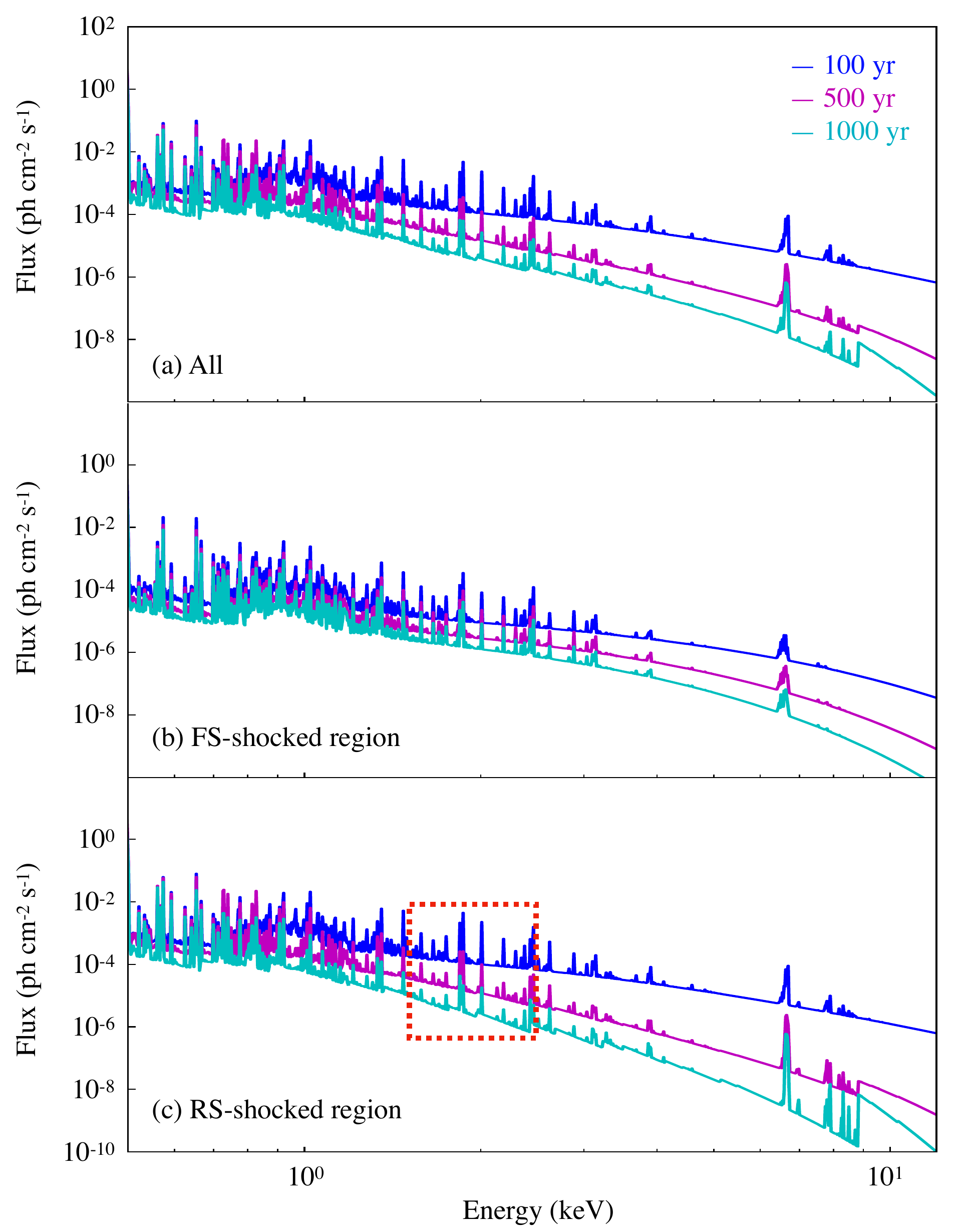}
	\includegraphics[width=80mm]{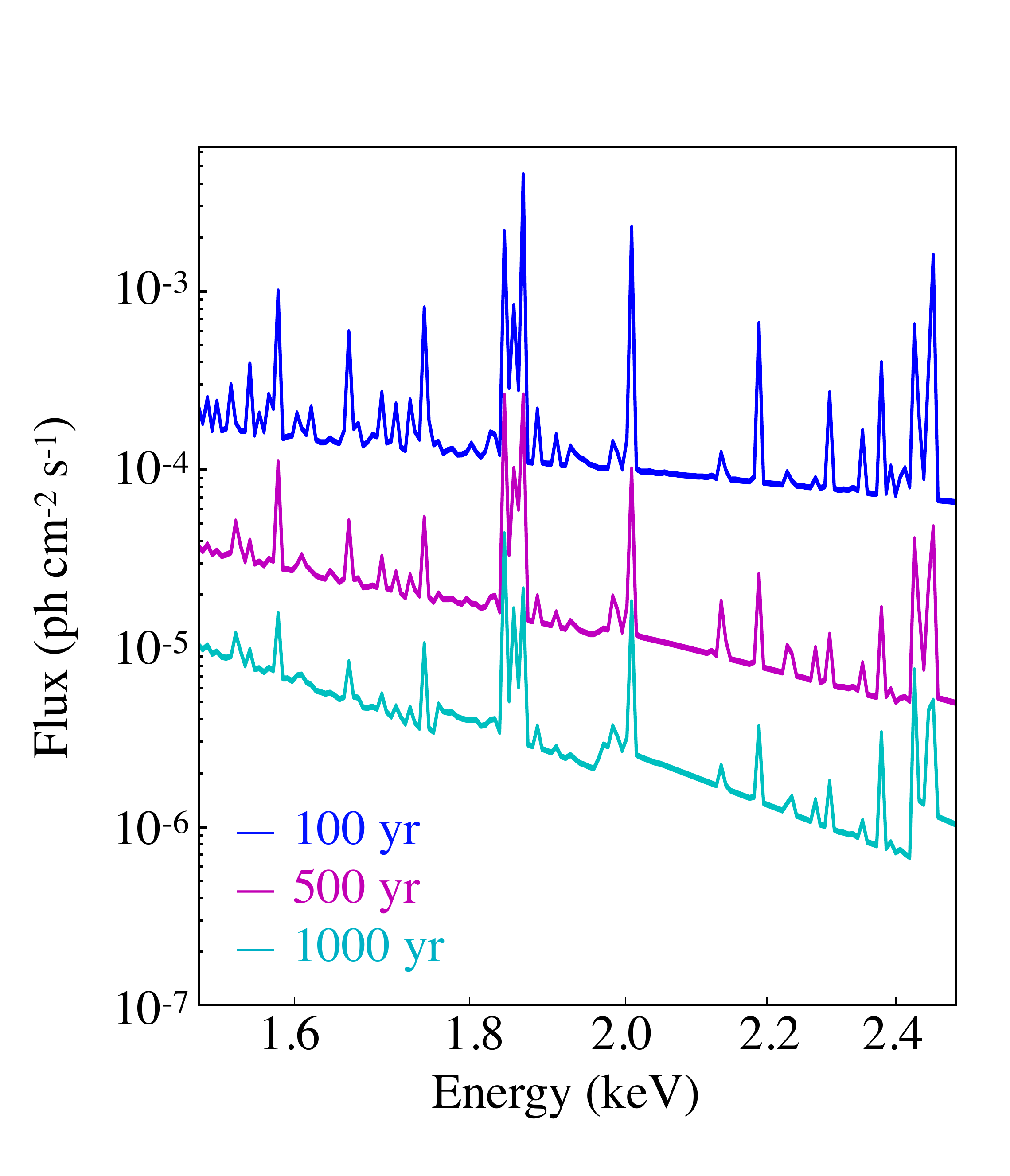}
	\caption{Time evolution of X-ray spectra integrated over the whole SNR (a), shocked CSM  (b), and shocked ejecta (c) at 100 yr, 500 yr and 1000 yr (left). The spectra in the ejecta region from 1.5 keV to 2.5 keV enclosed by the red square in the left panel is zoomed up and shown in the right panel.}
	\label{fig:spec_all_fs_rs}
\end{figure*}

X-ray spectra generated at the three different locations as in Fig.~\ref{fig:timeevol_te} at 1000 years are presented in Fig.~\ref{fig:spec_5}. 
The continuum structure of the spectra shows that the electron temperature is lower inside the shocked ejecta than the other regions, consistent with what we have seen in Fig.~\ref{fig:timeevol_te}. 
The radiative recombination continuum (RRC) structures of O and Ne appear around 0.65~keV and 1.2~keV in the spectrum from the shocked ejecta (red line). 
This indicates that the recombining process is dominant in the fast-expanding ejecta as we discussed in the previous section. 
The continuum spectra from the outer regions (blue and magenta) are remarkably similar to each other due to their similar electron temperatures.
However, their ionization states are obviously quite different from the line emissivities and ratios, attributable to their different ionization histories. 

Our framework can synthesize broadband spectra integrated over arbitrary regions at arbitrary ages.
As described in Section \ref{sec:time_evo}, emissions from the CSM and ISM are due to the plasma shocked by the FS and emissions from the ejecta are due to the RS.
We will refer to these two components as the emissions from the FS and RS heated regions, respectively.
For example, spectra integrated over the whole SNR, the FS-heated region, and the RS-heated region at 100 yr, 500 yr and 1000 yr are shown in Fig.~\ref{fig:spec_all_fs_rs}. 
In general, the continuum component over 1~keV becomes softer as time evolves due to a decreasing average electron temperature by the expansion of the SNR. 
A RRC structure from Fe (9~keV) is identified again in the ejecta region at 500 and 1000 years. 
\textbf{In this study, unless otherwise mentioned, the X-ray emission spectra are volume-integrated over regions in specific ranges along the radial direction from the explosion center. The volume emissivity in each Lagrangian mass layer is computed using the local thermal and ionization information at any given age.} 
%In this study, the spectra are calculated by integrating the emissions over a specific range along the radial direction, taking into account the volume of the sphere. The emissions of each layer is calculated from the emissions per unit volume calculated from the 1D model and the volume of each layer of the sphere.} 
This treatment is suitable for studying the global trend of the time evolution of the ejecta and ISM. Note that integration on the line-of-sight, instead of the radial profile, after the three-dimensional reconstruction of an SNR is also possible with our framework \textbf{(see Appendix)}, and will be necessary for detailed data analysis with spatially resolved spectroscopy.

%====== Temperatures for comparing with observations =====
\subsection{Comparison with $T_e$ and $T_Z$ from observations}
 
\begin{figure}[t]
	\centering
	\includegraphics[width=70mm]{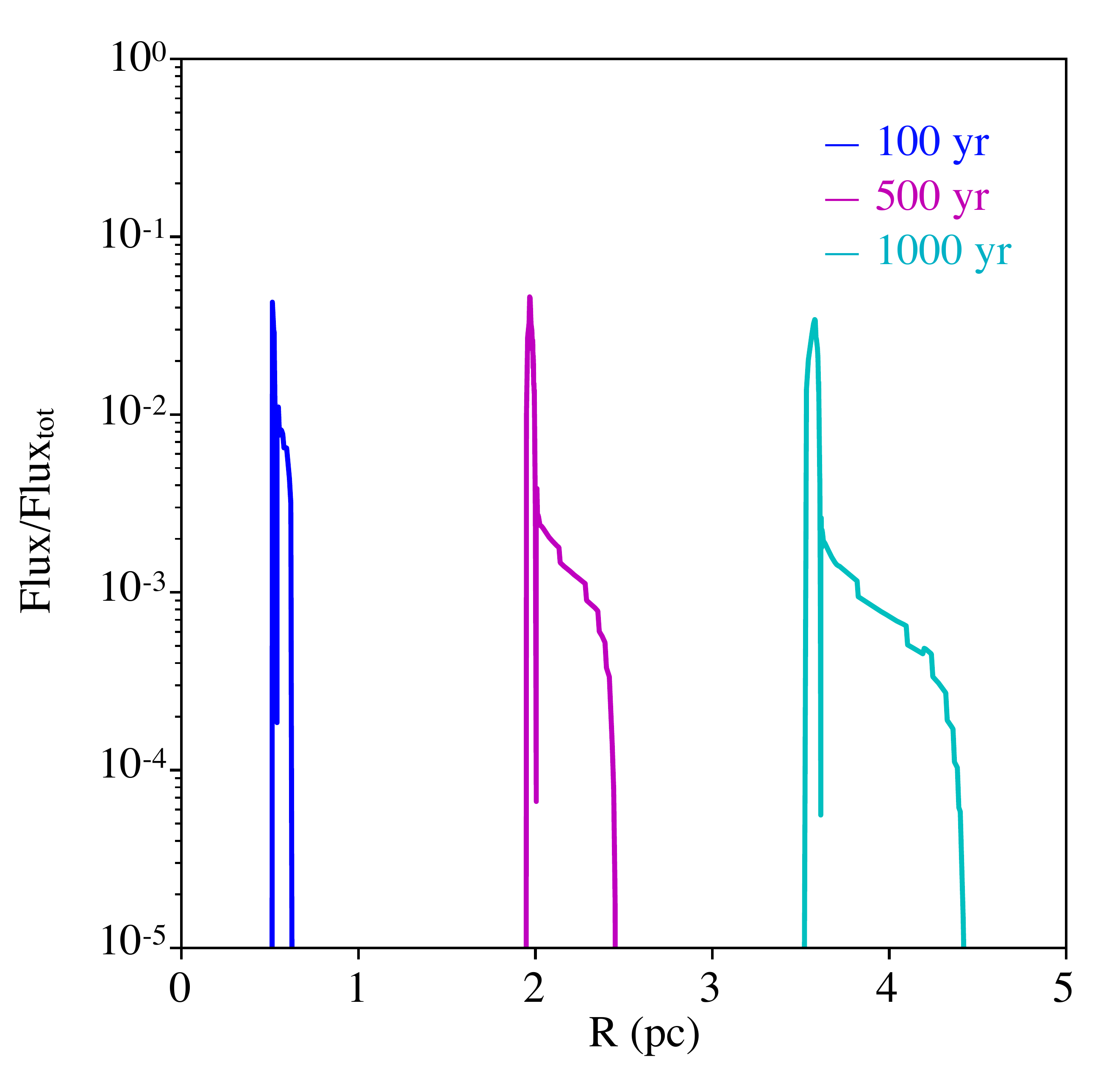}
	\caption{Snapshots of the radial profile of the fractional X-ray flux (normalized by the total fluxes in the 0.5--10 keV band) at the age of 100, 500 and 1000 years respectively. At each epoch, the locations of the FS and RS roughly correspond to the right and left boundaries of the profiles, enclosing the regions of shocked ejecta and shocked CSM. An abrupt drop of flux can be seen at the CD, which is a numerical artifact but does not affect our results in any significant way.}
	\label{fig:flux_temp}
\end{figure}

Observationally, the electron and ionization temperatures are inferred from spatially integrated spectral data from telescopes onboard satellites. 
Therefore, in order to compare our calculated temperatures from simulations with those from observations, we first need to provide a temperature  appropriately averaged over its spatial profile at any given age.
In this study, we use an average temperature weighted by the X-ray flux from 0.5~keV to 10~keV at each fluid element, which is a commonly observed energy band in X-ray observational analysis.
Fig.~\ref{fig:flux_temp} shows the fractional X-ray flux in each spatial layer at 100 yr, 500 yr and 1000 yr, which  is normalized by the total flux.

\begin{figure}[t]
	\centering
	\includegraphics[width=85mm]{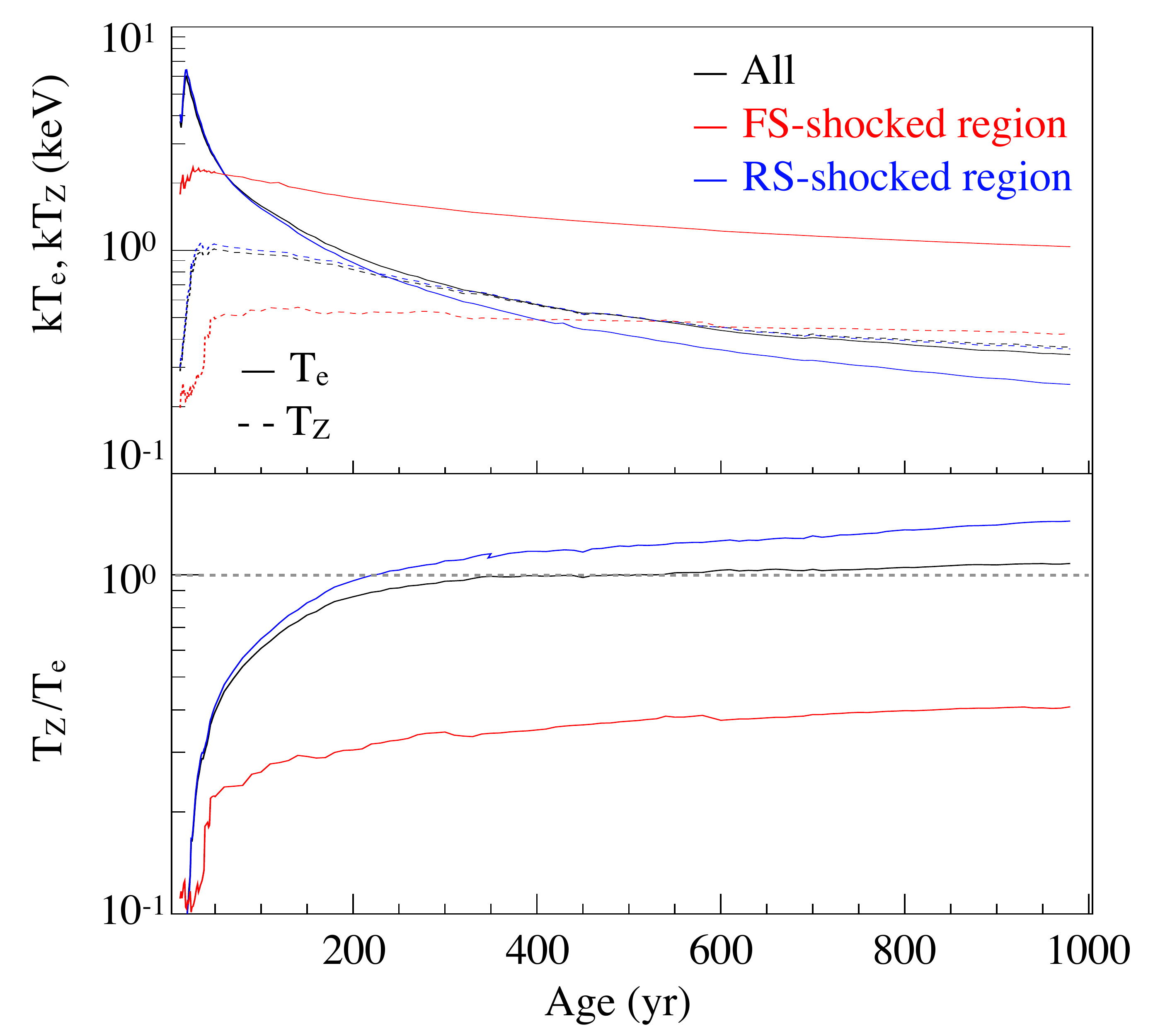}
	\caption{Upper panel: time evolution of electron temperature $T_e$ (solid) and ionization temperature $T_Z$ (dotted) averaged over the entire SNR (black), FS-heated CSM (red) and RS-heated ejecta (blue). The averages are weighted by the X-ray flux distribution shown in Fig.~\ref{fig:flux_temp}. Bottom panel: ratio of $T_Z/T_e$ for the same three regions.}
	\label{fig:tetz_all}
\end{figure}

The time evolution of the $T_e$ and $T_Z$ averaged over the entire SNR, the RS-heated region, and the FS-heated region respectively are presented in the top panel of Fig.~\ref{fig:tetz_all}.
These weighted averages allow us to emphasize regions that are the most representative of the X-ray emitting plasma in the remnant.
The bottom panel shows the ratio of $T_Z/T_e$ which provides a measure for the ionization state of the plasma.
The time evolution of the temperatures and their ratios averaged over the entire SNR are similar to those found in the shocked ejecta, since the total X-ray flux from the remnant has been dominated by the shocked ejecta throughout the first 1000 years.
This is evident in Fig.~\ref{fig:flux_temp} where we can see that the fractional X-ray flux is significantly higher in the shocked ejecta than behind the FS.

The overall time evolution is in general similar to what is shown in Fig.~\ref{fig:timeevol_te} for individual fluid elements, where both $T_e$ and $T_Z$ experience an initial sharp rise followed by a gradual decrease with time, for the same physical reasons explained in Section~\ref{sec:time_evo}.  
We can see that both $T_e$ and $T_Z$ for the shocked ejecta are initially higher and decrease faster with time afterward than in the shocked CSM.
This can be explained by the following. 
A strong RS was driven into the dense outer ejecta immediately after the explosion as the ejecta runs into the dense inner wind material at high speeds. 
The dense shocked ejecta contributes to a fast Coulomb heating of the electrons quickly toward equilibrium with the proton temperature, and hence the sharp boost of $T_e$ initially.
However, the RS had to climb up a steep ejecta density gradient in the early phase, leading to its fast deceleration shortly after the explosion, whereas the FS is propagating with a much milder deceleration in a wind-like CSM with the density decreasing as $r^{-2}$. 
The RS becomes weak quickly (see Fig.~\ref{fig:timeevol}), and a rapid decline of $T_e$ in the shocked ejecta follows due to the fast expansion of the ejecta.
As the SNR continues expanding outward into the tenuous wind-like CSM, the RS stays weak throughout the first 1000~yr because the accumulated mass swept up by the FS is still small compared to the ejecta mass. 
On the other hand, the FS which experiences only moderate deceleration continues to inject hot plasma into its downstream region.
Here, $T_e$ does not reach equilibrium with the protons so quickly as in the ejecta due to the low density in the shocked wind. 
Coulomb interaction with the hot protons and ions can hence continue to heat up the electrons effectively in the shocked CSM which helps slow down the cooling due to adiabatic expansion. 

A similar tendency can be observed in $T_Z$, but in a milder way since the ionization/recombination are typically slower than the heating/cooling processes in SNR plasma. 
As shown in the bottom panel of Fig.~\ref{fig:tetz_all}, while $T_Z/T_e$ stays below $1$ for the FS-heated gas, it evolves from $<1$ (under-ionized) to $>1$ (over-ionized) for the plasma in the shocked ejecta.
The fast cooling of the plasma in the ejecta proceeds much faster than the recombination of the free electrons with the ions. 
The decrease of $T_Z$ with time hence lags behind the electron cooling, and at an age around 300 yr the ionization fractions become higher than their CIE values, i.e., an over-ionization state with $T_Z/T_e > 1$. 

%=== Dependence of temperatures on surrounding environments === 
\section{Results}\label{sec:results}
Our simulation framework enables us to investigate the effect of different CSM environments on the time evolution of ionization states in SNRs through a parametric study.
In this section, we perform a survey on a two-dimensional parameter space defined by ($n_{\rm ISM}$, $\dot{M}_{\rm wind}$), i.e., the ambient ISM density and the pre-SN mass-loss rate of the progenitor. 
As described in Section \ref{sec:init_cond}, we determined the position of the physical connection between a wind-like CSM and an outer homogeneous ISM by considering either the pressure balance or the time duration of the mass loss. 
To facilitate a parametric study, we first fix the mass-loss rate at a fiducial value of $1 \times 10^{-5}$~M$_\odot$~yr$^{-1}$ and vary $n_{\rm ISM}$ from 0.1 to 30 cm$^{-3}$ in the first part (Section \ref{chan_ism}), and then we fix $n_{\rm ISM}$ at 0.2 cm$^{-3}$ and vary $\dot{M}_{\rm wind}$ from $1 \times 10^{-6}$ to $1 \times 10^{-4}$~M$_\odot$~yr$^{-1}$ in the second part (Section \ref{chan_dmdt}).
The parameter values of the wind velocity, mass of the ejecta from a CCSN (model \textit{s25D}), initial kinetic energy, and mass of the CSM are set to $20$~km~s$^{-1}$, $12.3$~M$_\odot$, $1.2 \times 10^{51}$~erg, and $8$~M$_\odot$, respectively, which are the same as in Section 3.
We follow the evolution of the resulted remnant up to an age of $5\times10^4$~years after the explosion, and study the dependence of $T_e(t)$, $T_Z(t)$ and $(T_Z/T_e)(t)$ on the CSM models.

\begin{figure}[t]
	\begin{center}
	\includegraphics[width=80mm]{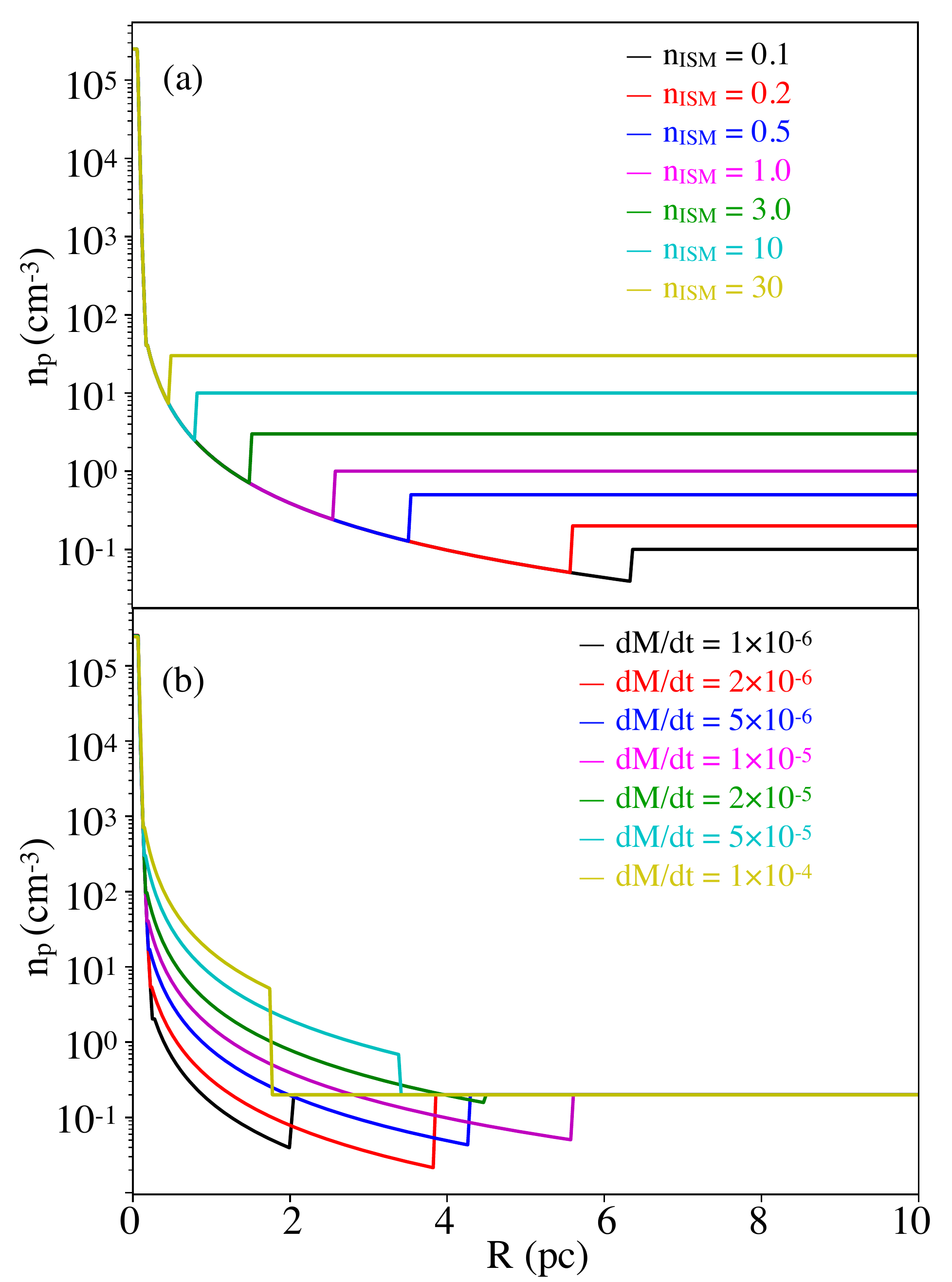}
	\end{center}
	\caption{Upper panel: initial density profiles for models with a variation of ISM densities $n_{\rm ISM} = 0.1$, 0.2, 0.5, 1.0, 3.0, 10 and 30~cm$^{-3}$, and a fixed mass-loss rate $\dot{M}_{\rm wind} = 1 \times 10^{-5}$~M$_{\odot}$~yr$^{-1}$.  
	Bottom panel: initial density profiles for models with a variation of mass-loss rates $\dot{M}_{\rm wind} = 1\times10^{-6}$, $2\times10^{-6}$, $5\times10^{-6}$, $1\times10^{-5}$, $2\times10^{-5}$, $5\times10^{-5}$ and $1\times10^{-4}$~M$_{\odot}$~yr$^{-1}$, and a fixed ISM density $n_{\rm ISM} = 0.2$~cm$^{-3}$.
	Wind velocity $v_{\rm wind} = 20$~km~s$^{-1}$ and total mass loss $M_{\rm wind} = 8$~M$_{\odot}$ in all cases. The leftmost structure is the density profile of the SN ejecta with a mass $M_{\rm ej} = 12.3$~M$_\odot$.}
	\label{fig:s25_various_init}
\end{figure}

%========== Dependence on the ISM density ================
\subsection{Dependence on the ISM density}\label{chan_ism}
The upper panel of Fig.~\ref{fig:s25_various_init} shows the initial density profile of our models, with 7 values chosen for $n_{\rm ISM} =$~0.1, 0.2, 0.5, 1.0, 3.0, 10, and 30 $\rm cm^{-3}$.
The mass-loss rate is fixed at $\rm1 \times 10^{-5}$~M$_{\odot}$~yr$^{-1}$.

\begin{figure*}[tb]
	\begin{center}
	\includegraphics[width=180mm]{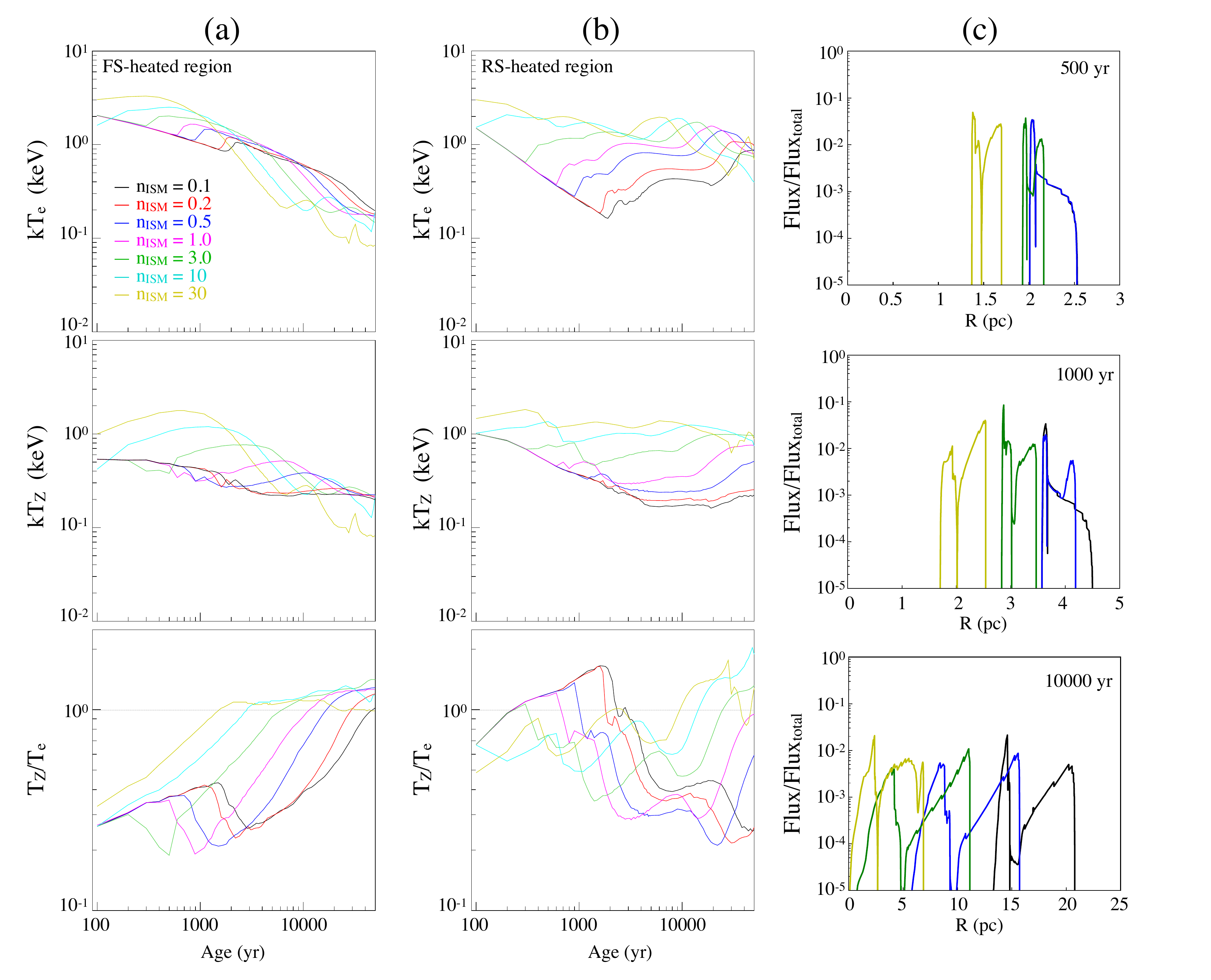}
	\end{center}
	\caption{Time evolution of $kT_e$ (top), $kT_Z$ (middle) and $T_Z/T_e$ (bottom) in the FS-heated (Panel (a)) and RS-heated (Panel (b)) regions for models with a fixed $\dot{M}_{\rm wind} = 1 \times 10^{-5}$~M$_{\odot}$~yr$^{-1}$ but with $n_{\rm ISM}$ varied from 0.1 to 30~cm$^{-3}$. Panel (c) shows snapshots of spatial profile of the fractional X-ray flux for $n_{\rm ISM} =$~0.1, 0.5, 3, and 30 $\rm cm^{-3}$  at ages of 500, 1000, and 10000 years respectively.}
	\label{fig:s25_ism}
\end{figure*}
Results of our models with different ISM densities are shown in Fig.~\ref{fig:s25_ism} from 100~yr to $5\times10^4$~yr.
As in the previous section, the temperatures are averaged over the FS-heated and RS-heated regions weighted by the fractional X-ray flux in each fluid element. 
Regardless of the ISM densities, $T_e$ and $T_Z$ follow the same evolution until the FS reaches the ISM.
The FS enters from the wind to the ISM region at a different age depending on the ISM density, at which a density jump causes a small boost of $T_e$ in the FS-heated region.
From this point on, the X-ray flux from the FS side becomes increasingly dominated by the shocked ISM material rather than the wind shocked in the past (see panel (c) in Fig.~\ref{fig:s25_ism}), as the mass swept up by the FS in the ISM becomes higher than the total mass in the wind. 
As the heating begins to weaken due to the deceleration of the FS, $T_p$ and $T_e$ decrease and Coulomb heating becomes less effective behind the FS. 
This effect happens at an earlier time for higher $n_{\rm ISM}$ as shown in panel (a) of Fig.~\ref{fig:s25_ism} for obvious reason. 

The FS becomes radiative eventually (e.g., at around $10^4$~years for the cases with the highest ISM densities), causing a sudden slow down of the shock.
The plasma immediately behind the FS then drops below X-ray emitting temperatures quickly, and the $T_e$ and $T_Z$ are now mainly determined by the hot plasma further downstream where radiative cooling has not yet become important. 
We can see fluctuations in both $T_e$ and $T_Z$ after the radiative transition, which are most visible for $n_{\rm ISM} = 10$ and $30$~cm$^{-3}$.
These are due to intermittent compressions of the downstream plasma as they collide with the slow and cold dense shell of gas immediately behind the radiative FS. 
Regardless of the details above, $T_Z/T_e$ gradually approaches unity as the plasma works its way toward a CIE state.
The higher $n_{\rm ISM}$ is, the shorter it takes for the plasma to reach ionization equilibrium. 

On the RS side, the plasma is found to be in an over-ionized state in the early phase for cases with low $n_{\rm ISM}$ due to a longer recombination timescale than the dynamical timescale as already explained in the previous section. After the entrance of the FS into the ISM, however, a revival of the RS happens, and the expansion of the shocked ejecta slows down. $T_e$ hence starts increasing, with $T_Z$ showing a similar evolution as well but in a milder way due to the relatively long ionization timescale in the now expanded and tenuous ejecta. With efficient heating, $T_Z/T_e$ has now dropped below $1$ and the ejecta becomes under-ionized.

For a high $n_{\rm ISM}$, $T_Z/T_e$ can increase again as the SNR becomes radiative for the following reason. The pressure in the shocked ISM drops quickly due to the radiative loss, and as a result the expansion of the shocked ejecta accelerates outward, causing a drop in $T_e$. We can see that $T_Z$ does not drop immediately with $T_e$ as the recombination lags behind the adiabatic cooling. An overall increase in $T_Z/T_e$ hence results. For $n_{\rm ISM} = 10$ and $30$~cm$^{-3}$, the ejecta becomes over-ionized again after $\sim 10^4$~years.  From our results, it is interesting to observe that the evolution of the ionization state in SNR plasma can be far from monotonic, and is found to be sensitive to the surrounding environment of the SNR even without invoking non-trivial density structures such as the presence of clumpy molecular clouds. The case for $n_{\rm ISM} = 30$~cm$^{-3}$ can however be used to mimic situations when a CCSN explodes inside a dense cloud-like environment. 

%=========================================================
\subsection{Dependence on the pre-SN mass-loss rate}\label{chan_dmdt}
\begin{figure*}[tb]
	\begin{center}
	\includegraphics[width=180mm]{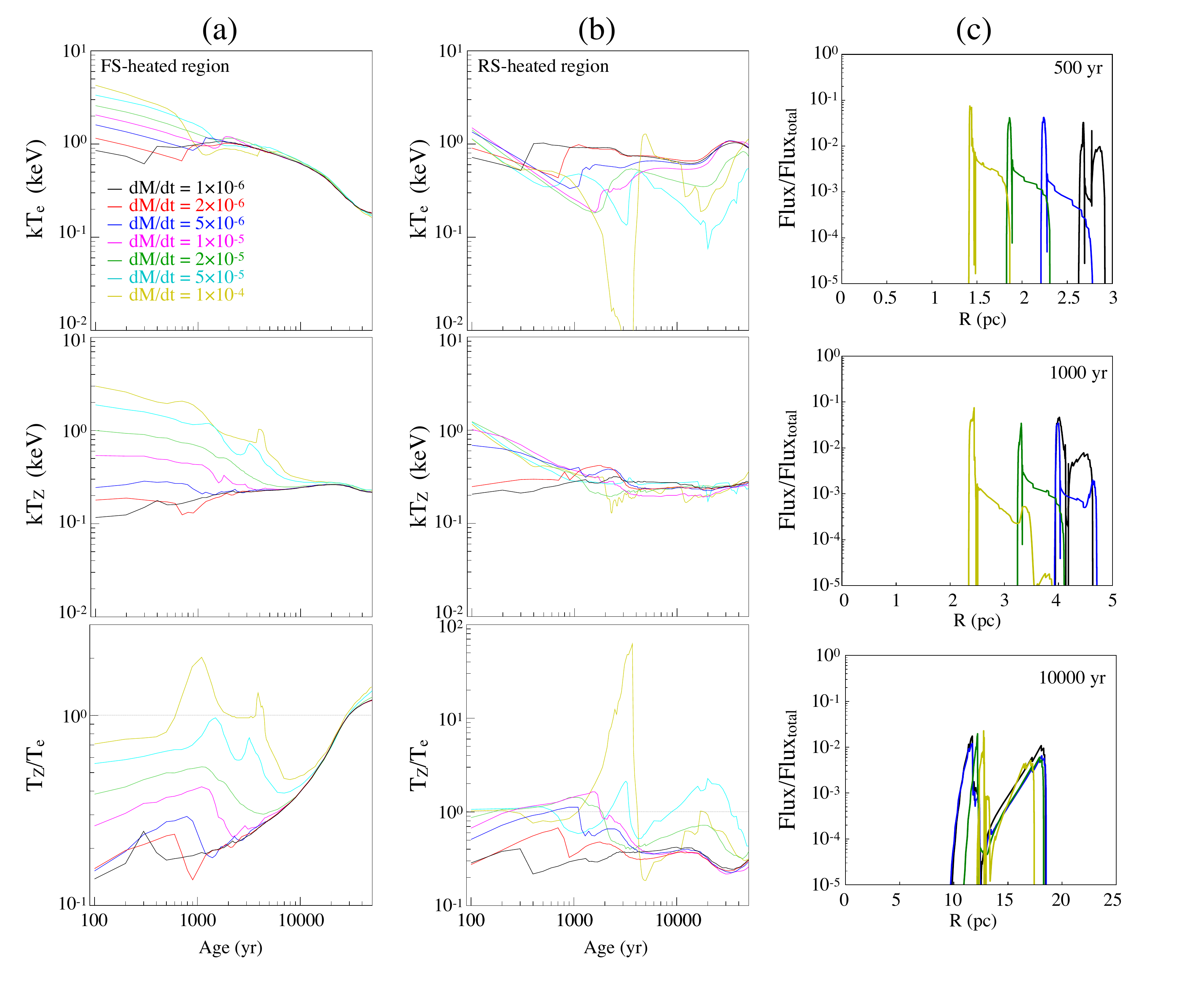}
	\end{center}
	\caption{
	Same as Fig. \ref{fig:s25_ism} but with $n_{\rm ISM}$ fixed at 0.2 cm$^{-3}$ and $\dot{M}_{\rm wind}$ varied from $1 \times 10^{-6}$ to $1 \times 10^{-4}$~M$_\odot$~yr$^{-1}$. Panel (c) shows snapshots of spatial profile of the fractional X-ray flux for $\dot{M}_{\rm wind} =$~$1 \times 10^{-6}$, $5 \times 10^{-6}$, $2 \times 10^{-5}$, and $1 \times 10^{-4}$~M$_\odot$~yr$^{-1}$ at ages of 500, 1000, and 10000 years respectively.
	}
	\label{fig:s25_dmdt}
\end{figure*}

Using the CSM models shown in panel (b) of Fig.~\ref{fig:s25_various_init}, we investigate the effect of pre-SN mass-loss rate on the long-term evolution of ionization state in SNRs. The results are shown in Fig.~\ref{fig:s25_dmdt}. 

For the FS-heated region (panel (a)), the behavior of $T_e$ is qualitatively similar to what we have observed and discussed already in Section \ref{chan_ism}. 
In the early phase, the evolution is determined by the shocked wind, where $T_e$ decreases with time via adiabatic cooling but hindered in the meantime by the Coulomb heating from the ions.
The absolute values of $T_e$ are higher for cases with faster mass losses, simply because of a higher density in the wind and hence a more efficient ``boost'' by Coulomb interaction behind the FS.
After the entrance of the FS into the ISM region with $n_{\rm ISM} = 0.2$~cm$^{-3}$ (which happens at different ages for different $\dot{M}_{\rm wind}$ in correspondence with the transition radii shown in Fig.~\ref{fig:s25_various_init}), $T_e$ converge back to a common trend for all models which is basically identical to the red curve in the upper panel of Fig.~\ref{fig:s25_ism} (a). 
The reason is that the total mass in the wind $M_{\rm wind}$, which is identical among the models, is small compared to $M_{\rm ej}$, so that the interaction of the SNR with the wind does not bear a strong importance to the dynamical evolution. 

$T_Z$ on the other hand shows a more interesting dependence on $\dot{M}_{\rm wind}$, where we see a $T_Z$ rising throughout the SNR's life up to $5\times10^4$~years for a $\dot{M}_{\rm wind} =$~$1\times10^{-6}$~M$_{\odot}$~yr$^{-1}$, whereas it drops almost monotonically for a $\dot{M}_{\rm wind} > 5\times10^{-5}$~M$_{\odot}$~yr$^{-1}$. 
Collisional ionization proceeds slowly towards CIE in a tenuous wind and ISM, as we can see in the evolution of $T_Z/T_e$ shown in the bottom panel. 
For cases with high $\dot{M}_{\rm wind}$, the interaction with a dense wind in the vicinity of the ejecta shortly after the explosion leads to a quick rise of $T_Z$ in the very beginning.
As the wind density decreases rapidly as $r^{-2} \sim t^{-2}$ considering an approximately free expansion, the ionization timescale also increases quickly and the averaged $T_Z$ decreases.
In particular, for the case of $\dot{M}_{\rm wind} = 1\times10^{-4}$~M$_{\odot}$~yr$^{-1}$, the FS basically ``breaks out'' from the dense wind into the tenuous ISM at a radius of about 2.7~pc. 
This break out of the shock from a dense gas into a low-density medium leads to a moment of accelerated expansion of the remnant, which can result in a transient state of over-ionization in the plasma. 
This phenomenon is especially prominent in the shocked ejecta as we will discuss in the following.

For the RS-heated region (panel (b)), again the cases for $\dot{M}_{\rm wind} = 1\times10^{-6}$, $5\times10^{-6}$ and $1\times10^{-5}$~M$_{\odot}$~yr$^{-1}$ show results which are qualitatively similar to the previous section. 
The two cases with the highest mass-loss rates however display some characteristically different behaviors. 
In particular, the case with $\dot{M}_{\rm wind} = 1\times10^{-4}$~M$_{\odot}$~yr$^{-1}$ predicts a significant drop of $T_e$ at ages from about 1500 to 3000~years.
As mentioned above, this is attributed to the break out of the expanding shock front from the dense wind into the tenuous ISM region. 
First of all, the shocked ejecta experiences a fast adiabatic cooling from the accelerated expansion of the SNR.
The dense ejecta material shocked shortly after the explosion cools down to a $T_e \sim$ a few $10^5$~K at an age $\sim 2000$~years, at which the radiative cooling timescale decreases dramatically, creating a cold dense shell immediately behind the CD.
The averaged $T_e$ in the shocked ejecta thus drops rapidly further.
$T_Z$ does not show any drastic decline as the recombination timescale is much longer than the cooling timescale, leading to a phase in which the ejecta becomes over-ionized.
However, we note that from about 3000~years, the averaged $T_e$ has already dropped to about 0.01~keV so that the ejecta is no longer effective in emitting X-rays strongly, meaning that spectral signatures from the recombining plasma cannot be picked up by X-ray observations anymore. 
Up to this point, the RS stays weak and sits close to the CD.

At around 4000~years old, the total swept-up mass behind the FS becomes comparable to $M_{\rm ej}$, and a strong RS is driven into the ejecta as a result, reheating the plasma to X-ray emitting temperatures. 
The ejecta becomes under-ionized again and gradually approaches CIE with time. 

The case with $\dot{M}_{\rm wind} = 5\times10^{-5}$~M$_{\odot}$~yr$^{-1}$ sits in the middle between the two extremes.
The SNR break out does not happen as dramatically as in the densest wind model, but significant cooling of the plasma in the ejecta still occurs from 2000 to  20000~years.
One crucial difference here is that the plasma achieves an over-ionized state over a relatively long period of time while keeping a $kT_e$ from 0.1 to a few 0.1~keV. 
This means that X-ray spectroscopy should be able to detect signatures from the recombining plasma for a CCSNR surrounded by a moderately dense wind.

According to our results, recombining plasma cannot be realized in a SNR ejecta either when the mass-loss rate is too low or too high.
When the mass-loss rate is low, $T_z$ remains low due to the long ionization timescale in the thin gas where the density becomes small by the fast expansion.
In the case of high mass-loss rates, $T_e$ decreases by the break out of the expanding shock and radiative cooling makes the $T_e$ too low to emit X-rays. 
Observationally, such break out phenomenon and the accompanied plasma rarefaction have been suggested to be the case for SNR W44 \citep{uchida2012} and W49B \citep{miceli2010, zhou2011, yamaguchi2018}.

%=========== Comparison with the observation =============
\subsection{Comparison with observations}

\begin{deluxetable*}{cccccccccc}
	%\tablenum{1}
	\tablecaption{Observation results.\label{tab:observation_data}}
	\tablewidth{0pt}
	\tablehead{
		\colhead{Object} & \colhead{Age} & \colhead{Region} & \colhead{$n_{e}t$} & \colhead{$kT_{init}$} & \colhead{$kT_{e}$} & \colhead{$kT_{z}$} & \colhead{Reference} \\
		\colhead{} & \colhead{(yr)} & \nocolhead{} & \colhead{(10$^{11}$ cm$^{-3}$ s)} & \colhead{(keV)} & \colhead{(keV)} & \colhead{(keV)} & \colhead{No.}
	}
	\startdata
W49B       & 2900--6000   & Whole        &  4.4 &  5.9 & 0.94$^{+0.02}_{-0.02}$    & 1.878$\pm$0.04            & 1 \\
IC~443     & 3000--30000  & SE (Hot)     &  4.2 &  5.0 & 0.54$^{+0.03}_{-0.01}$    & 1.49$\pm$0.04             & 2 \\
           &              & SE (Cold)    &  4.2 &  5.0 & 0.19$^{+0.01}_{-0.01}$    & 1.17$\pm$0.03             &  \\
G359.1-0.5 & 10000--70000 & West         &  4.6 &  4.0 & 0.13$\pm$0.02             & 1.019$^{+0.1}_{-0.9}$     & 3 \\
W28        & 33000--36000 & Reg 4 (Hot)  &  3.7 &  3.8 & 0.520$^{+0.001}_{-0.002}$ & 1.53$^{+0.04}_{-0.08}$    & 4 \\
           &              & Reg 4 (Cold) & 10.1 &  3.8 & 0.216$^{+0.001}_{-0.005}$ & 0.269$^{+0.004}_{-0.052}$ &  \\
W44        & 10000--20000 & Whole        &  3.5 & 0.93 & 0.37$\pm$0.01             & 0.38$\pm$0.01             & 5 \\ 
G346.6-0.2 & 14000--16000 & Whole        &  4.8 &  5.0 & 0.30$^{+0.03}_{-0.01}$    & 1.20$^{+0.06}_{-0.15}$    & 6 \\ 
3C391      & 7400--8400   & Whole        & 14.0 &  1.8 & 0.495$\pm$0.015           & 0.49$^{+0.34}_{-0.01}$    & 7 \\
CTB~37A    & 6000--30000  & Whole        & 13.0 &  5.0 & 0.490$^{+0.09}_{-0.06}$   & 0.49$^{+0.39}_{-0.05}$    & 8 \\
G290.1-0.8 & 10000--20000 & Centre       & 12.2 &  1.7 & 0.45$^{+0.02}_{-0.01}$    & 0.45$^{+0.35}_{-0.01}$    & 9 \\ 
N49        & 4800         & Whole        &  7.0 & 11.0 & 0.62$\pm$0.01             & 1.21$^{+0.01}_{-0.04}$    & 10 \\ 
Kes~17     & 2000--40000  & Whole        & 16   &  5.0 & 0.73$\pm$0.03             & 0.87$^{+0.02}_{-0.05}$    & 11 \\ 
G166.0+4.3 & 24000        & NE           &  6.1 &  3.0 & 0.46$\pm$0.03             & 1.13$\pm$0.05             & 12 \\ 
3C400.2    & 14000--32000 & NE           &  1.8 & 3.15 & 0.71$^{+0.03}_{-0.02}$    & 2.1$^{+0.3}_{-0.2}$       & 13 \\ 
N132D      & 2500         & Whole        &  8.8 & $>$8 & 1.5$^{+0.1}_{-0.2}$       & 1.87$^{+0.2}_{-0.4}$      & 14 \\ 
HB~21      & 4800--15000  & Whole        &  3.2 & 0.58 & 0.17$^{+0.01}_{-0.02}$    & 0.31$^{+0.01}_{-0.16}$    & 15 \\ 
CTB~1      & 9000--44000  & SW           &  9.3 &  3.0 & 0.186$^{+0.008}_{-0.007}$ & 0.188$^{+0.005}_{-0.004}$ & 16 \\
Cas~A      & 341          & IME-rich     & 1.18 & -    & 1.7$\pm$0.1               & 0.89$^{+0.06}_{-0.08}$    & 17 \\
Kepler     & 417          & Whole        & 0.33 & -    & 0.37$\pm$0.01             & 0.236$^{+0.008}_{-0.007}$ & 18 \\ 
Tycho      & 449          & Whole        & 0.62 & -    & 0.70$^{+0.02}_{-0.01}$    & 0.270$^{+0.008}_{-0.009}$ & 19 \\ 
	\enddata
	\tablecomments{$^{1}$~\citet{smith1985, leahy2018, matsumuraD}, $^{2}$~\citet{petre1988, olbert2001, matsumura2017b}, $^{3}$~\citet{ohnishi2011, leahy2020, suzuki2020}, $^{4}$~\citet{velazquez2002, RandB2002, okon2018}, $^{5}$~\citet{smith1985, wolszczan1991, matsumuraD}, $^{6}$~\citet{yamaguchi2013}, $^{7}$~\citet{leahy2018, sato2014}, $^{8}$~\citet{abdollahi2020, sezer2011, yamauchi2014}, $^{9}$~\citet{slane2002, kamitsukasa2015}, $^{10}$~\citet{park2012, uchida2015}, $^{11}$~\citet{gelfand2013, washino2016}, $^{12}$~\citet{matsumura2017a}, $^{13}$~\citet{agrawal1983, ergin2017}, $^{14}$~\citet{VandD2011, bamba2018}, $^{15}$~\citet{leahy1987, lazendic2006, suzuki2018}, $^{16}$~\citet{KandH1991, craig1997, katsuragawa2018}, $^{17}$~\citet{fesen2006, katsuda2018}, $^{18}$~\citet{katsuda2015}, $^{19}$~\citet{katsuda2015}}
\end{deluxetable*}

\begin{figure*}[tb]
	\centering
	\includegraphics[width=180mm]{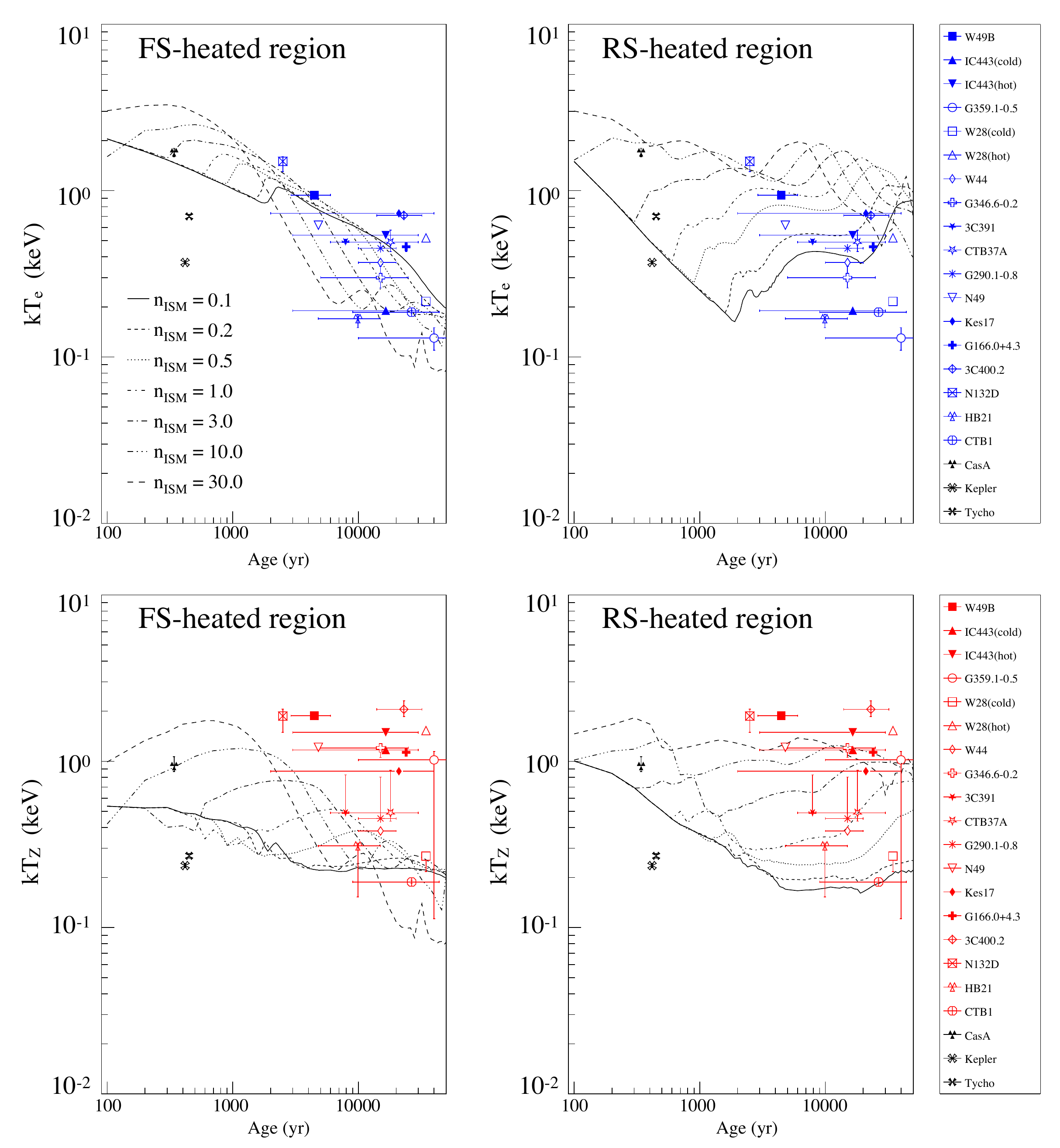}
	\caption{$T_e$ (top) and $T_Z$ (bottom) estimated from previous observations compared to the averaged temperatures extracted from our simulation models with $\rm n_{ISM}$ = 0.1, 0.2, 0.5, 1, 3, 10, and 30 cm$^{-3}$. Black points are temperatures of the young SNRs. Blue and red points mean the old SNRs. The mass-loss rate is fixed at $1\times10^{-5}$~M$_{\odot}$~yr$^{-1}$.}
	\label{fig:obs_s25_ism}
\end{figure*}

While it is beyond the purpose of this paper to develop best-fit models for any particular observed SNR, it is illustrative to provide a sanity check by comparing the results of our parametric study with previous observations of evolved SNRs for an examination of consistency.
\textbf{In order to quantitatively investigate the formation of overionized plasmas and to study individual objects, a multi-D approach would be necessary, as has been done in previous studies} \citep[e.g.][]{zhou2011, slavin2017, zhang2019}. This point is discussed in Section \ref{subsec:limitation}. However, our model is able to present global trends of the time evolution of SNRs, which will be useful to find an efficient way of two- or three-dimensional hydrodynamic modeling.

For a systematic comparison, we select nineteen objects whose X-ray spectra have been analyzed using NEI models.
The selected SNRs include all of old MM-SNRs in which evidences of recombining plasma have been found (W49B, IC~443, G359.1-0.5, W28, W44, G346.6-0.2, 3C 391, CTB~37A, G290.1-0.8, N49, Kes 17, G166.0+4.3, 3C400.2, N132D, HB~21, CTB~1), and three young SNRs (Cas~A, Kepler, and Tycho) for contrast.
The key properties of these SNRs are summarized in Table \ref{tab:observation_data}.
When the observational analysis was performed separately into multiple spatial regions, we adopted results only from regions where the signature of recombining plasma was clearly shown.
If multiple components are used for the X-ray model of the observed spectra such an ejecta and ISM component, we choose the ejecta component for our comparison.
Similar to the previous sections, the simulation results are presented as temperatures averaged over two regions, i.e., the shocked ejecta and the shocked ambient gas. 
The ion fractions from observations are calculated from the measured electron temperature, the initial ion fraction (or initial $T_e$) and ionization timescale $n_et$ using NEI models implemented in the \textit{PyAtomDB} toolkit which is a selection of utilities designed to interact with the AtomDB atomic database in the {\it Python} environment \citep[][]{atoms8030049}. 

Fig.~\ref{fig:obs_s25_ism} shows the electron and ionization temperatures obtained from the observations as data points, on which the model results are overlaid.  
The calculated range of $T_Z$ in the RS-heated region from our models appears to be in agreement with the observations of old SNRs, whereas the results for the FS-heated region disagree with the data as an overall under-prediction. 
This is consistent with the conclusions from most observational studies that the recombining plasma found is dominated by the ejecta components based on the chemical abundances inferred from the X-ray spectra.
On the other hand, several objects seem to have lower $T_e$ than those predicted by the simulations.

As we described Sec.~\ref{chan_dmdt} (Fig.~\ref{fig:s25_dmdt}), simulations with high mass-loss rates reproduce a low $T_e$, but $T_Z$ is under 0.3~keV that is lower than most of observations. The simulation results suggest that some objects are lying outside the limited parameter space surveyed in this study and imply that other physical processes are needed. It is possible that additional cooling mechanisms other than adiabatic and radiative processes are at work to achieve such low electron temperatures.

%=========================================================
%====================== Discussion ======================
\section{Discussion}\label{sec:disc}

\subsection{Possible formation scenarios of recombining plasma}

We have investigated the effect of different surrounding environments on the long-term evolution of ionization states in SNRs using a newly developed framework and found that even without invoking any non-trivial CSM structures, the calculated evolution paths are far from monotonic and highly dependent on the gas density.
In particular, SNR plasma does not simply evolve from an under-ionzied state gradually towards CIE with age as was commonly pictured before.
Within the scope of our models and chosen parameter space, there are mainly three possible situations that can produce an X-ray bright recombining plasma in an SNR:
\begin{enumerate}
    \item A RSG-like progenitor surrounded by a pre-SN wind with $\dot{M}_{\rm wind} \sim 10^{-5}$~M$_{\odot}$~yr$^{-1}$ and a tenuous ISM with $n_{\rm ISM} \lesssim 0.2$~cm$^{-3}$. In this case, recombining plasma can exist in the ejecta of young SNRs ($<$ a few 1000 years old) due to the rapid adiabatic cooling as the SNR expands in the extended RSG wind and breaks out into a tenuous ISM. This trend has also been seen in the multi-D study \citep{zhou2011}. As the RS revives when the FS has entered the ISM region and the SNR has swept up enough mass, the RS heats up the inner ejecta and the plasma reverts to an under-ionized state once more. 
    
    \item A progenitor with a pre-SN wind of $\dot{M}_{\rm wind} \sim 10^{-5}$~M$_{\odot}$~yr$^{-1}$ and a dense ISM of $>$3~cm$^{-3}$.  In this case, $T_e$ and $T_z$ are kept high because the high ISM density reduces the effect of expansion. When the pressure in the FS-heated region drops and the ejecta expands outward, the recombining plasma can be realized. In order for the ejecta to reach an over-ionized state at old ages (a few 1,000~years),  dense gas is needed in the surrounding.
    This result applies to a situation similar to the explosion of the progenitor inside a dense ISM such as a cloud-like environment.
    
    \item An SN progenitor which has experienced a phase of enhanced mass loss prior to core collapse. In this case, the SNR interacts with a dense wind after the explosion, but later on breaks out from the wind into a tenuous ISM. The break-out leads to a rapid cooling of the ejecta, possibly assisted by radiative cooling in the dense ejecta shocked early on. The recombination lags behind the rapid cooling, leading to an over-ionized state. The change to the over ionization can happen for SNRs older than a few 1000 years old. To realize recombining plasma in the SNR ejecta in a prolonged period, the mass-loss rate of the progenitor needs to be around $5\times10^{-5}$~M$_{\odot}$~yr$^{-1}$ so that $T_e$ does not decrease below X-ray emitting temperatures caused by strong radiative cooling. $\dot{M}_{\rm wind}$ cannot be too low neither since otherwise the ejecta density becomes too low too quickly by the fast expansion for ionization to proceed. 
\end{enumerate}

\subsection{Limitation of the one-dimensional model}\label{subsec:limitation}

As we have discussed above, although our models are quite successful in reproducing the range of $T_Z$ inferred from observations of evolved SNR, some objects show $T_e$ below our predictions. In addition to adiabatic and radiative cooling, other effective cooling mechanisms have been suggested in the literature for explaining the existence of recombining plasma in old SNRs. For example, interaction with dense gas clumps in a molecular cloud can cool down the plasma efficiently via thermal conduction and evaporation \citep[e.g.][]{WandL1991, seta2004, matsumura2017b, slavin2017, zhang2019, sashida2013, okon2020, okon2021}. Several SNRs displaying recombining plasma have also been found to emit gamma rays in the GeV band, which implies that the formation of over-ionized state has a link to interactions of molecular clouds and cosmic rays accelerated at an SNR \citep{suzuki2018}. In this context, other coolants may include efficient particle accelerations and turbulence generation\citep{bamba2005, uchiyama2007}, which are also not accounted for in this study.

In general, SNRs observed in X-rays exhibit asymmetric 3D structures, and many of them have small-to-medium scale heterogeneities in the surrounding environment (e.g., molecular clouds) \citep[e.g.][]{troja2006, keohane2007}. In addition, hydrodynamic instabilities that develop during the interaction of the remnant with CSM inhomogeneities need to be taken into account. For example, IC~443, an archetypal MM-SNR, has revealed elongated jet-like structures with magnesium-rich plasma in over-ionization \citep{greco2018}. It is difficult to consider such complicated 3D structures in our 1D model that assumes spherical symmetry. Thus, a 2D or 3D hydrodynamic model will be necessary to simulate complicated structures possibly emerging from interactions with localized cold molecular clumps or from hydrodynamic instabilities that can be suppressed in a 1D model. However, local physical processes---ionization/recombination and a variety of energy changes---should be the same both for a 1D and for a multi-D models. The physics implementation described in this work can be naturally incorporated into multi-D simulation for detailed comparisons with spatially resolved spectra of SNRs in the future. Furthermore, 1D models still have advantage in spatial resolution, which allows us to theoretically investigate effects of fine complicated structures, which can also be triggered by hydrodynamic instabilities and by multiple shock collisions.

\section{Concluding remarks}\label{sec:conc}

We have developed a new simulation framework to model the time evolution of the ionization state of the plasma in an SNR based on a one-dimensional hydrodynamics code and a fully time-dependent non-equilibrium ionization/recombination calculation.
This framework is capable of spectral synthesis in X-rays for comparisons with spatially resolved spectroscopic observations of SNRs at arbitrary ages.
We have calculated the long-term evolution of physical properties, up to 10$^{4}$~years, including the electron temperature and the ionization states for SNRs in different surrounding environments---the pre-SN stellar wind and the ISM.
The results show that a recombining plasma is realized when the electrons suffer rapid cooling via adiabatic expansion or strong soft X-ray radiation and when the recombination lags behind this fast change in the electron temperature.
Within the scope of our models and chosen parameters, such a situation can occur in cases of a fast expansion of an SNR in a spatially extended low-density wind, an SNR expanding in a dense ISM, and a break out of an SNR from a confined dense wind region into a tenuous ISM.
Particularly, the high ionization temperature of $\gtrsim$ 1 keV reproduced in the second case, an old SNR in a dense ISM environment, suggests that a dense could play an important role in a formation of recombining plasma, though an interaction with a small dense cloud should be investigated by multi-D hydrodynamic simulation.

Our simulation framework can be applied to the modeling of observation data with high-resolution spectroscopy.
For example, we can produce an extensive grid of models over a broad parameter space, which can be converted into table models readily usable by data analysis tools such as XSPEC. 
We anticipate the launches of next-generation X-ray observatories with micro-calorimeters such as XRISM and Athena, which will have at least 25 times better energy resolutions compared to conventional X-ray CCDs.
Our hydrodynamic simulation platform will be especially well suited for predicting and interpreting these future high-quality X-ray spectra from SNRs at any age of their evolutions.

\acknowledgments
M.K. and T.T. are supported by JSPS KAKENHI Grant No. JP18H05463.
S.H.L. is supported by JSPS KAKENHI Grant No. JP19K03913. 
H.O. is supported by JSPS KAKENHI Grant No. JP19H05185, JP19H01906 and JP18H05861.
T.T. is supported by JSPS KAKENHI Grant No. JP20H00153.
M.K., S.H.L., H.O. and T.T. acknowledge support by the World Premier International Research Center Initiative (WPI), MEXT, Japan.
We also thank an anonymous referee for helpful comments.
\vspace{5mm}
\software{ \href{http://wonka.physics.ncsu.edu/pub/VH-1/}{VH-1}, \href{https://root.cern}{ROOT}, \href{https://www.python.org/}{Python}, Numpy \citep{vanderwalt2011}, \href{http://atomdb.readthedocs.io/en/master/}{PyAtomDB}, Matplotlib \citep{hunter2007}.}

\appendix
\section*{Sky-projected image and spectra integrated along the line of sight}
\begin{figure*}[t]
	\centering
	\includegraphics[width=180mm]{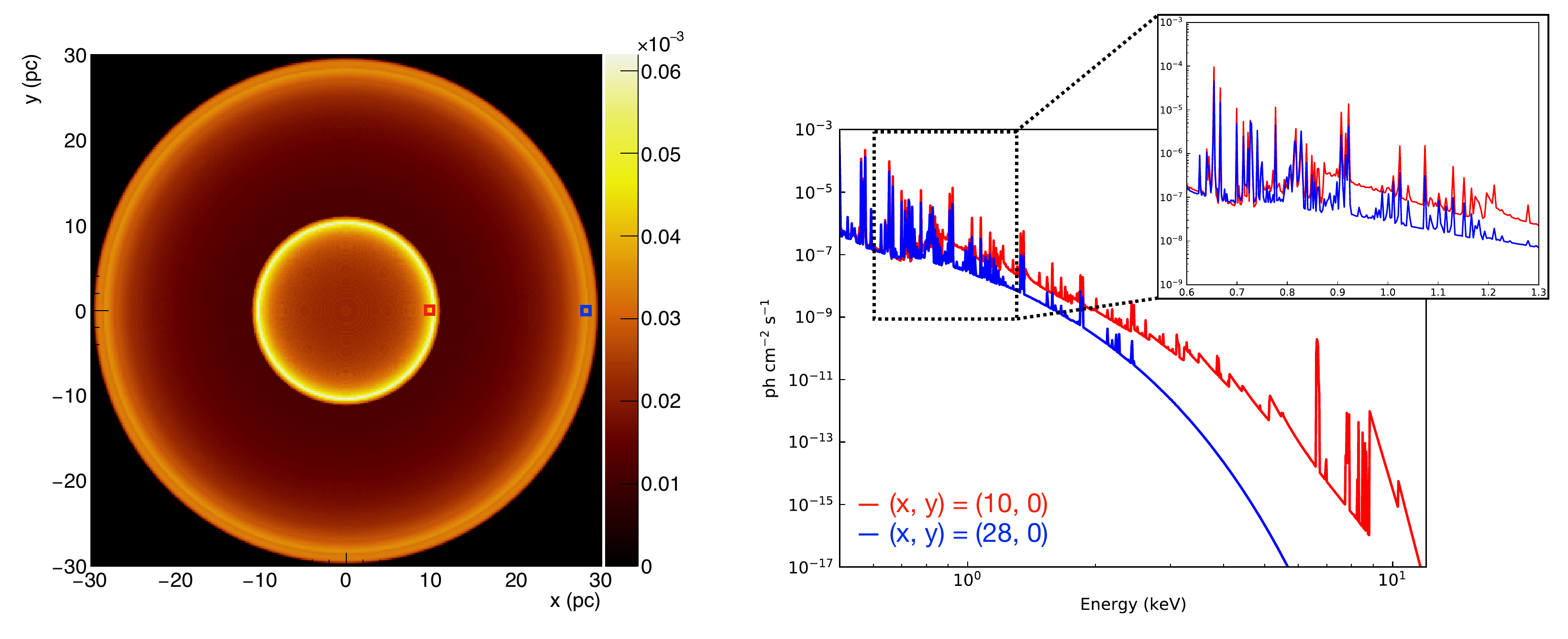}
	\caption{Sky-projected image (left) and spectra (right) of a SNR simulated with $5\times10^{-5}$~M$_{\odot}$~yr$^{-1}$ and $\rm n_{ISM}$ = 0.2~cm$^{-3}$ at 3$\times$1-$^4$~yr. Spectra are integrated along the line of sight in the region of 1~pc$\times$1~pc, which shown in red and blue lines in the sky-projected image.}
	\label{fig:apen}
\end{figure*}

Our code can provide Sky-projected images and spectra integrated along the line of sight. Figure \ref{fig:apen} shows an example of sky-projected image and spectra at 3$\times$10$^4$~yr in the case with $\dot{M}_{\rm wind}=5\times10^{-5}$~M$_{\odot}$~yr$^{-1}$ and $\rm n_{ISM}$ = 0.2~cm$^{-3}$. The X-ray emissions in the energy band of 0.5--12~keV are used for synthesizing the image. RRC structures appear in the X-ray spectrum extracted from a region around the contact discontinuity at (x, y) = (10, 0) pc (red square in the sky-projected image), while the structure disappear in the spectrum around the FS at (x, y) = (28, 0) pc (blue square in the sky-projected image).

\bibliography{mybibfile}{}
\bibliographystyle{aasjournal}

\end{document}